\documentclass[reprint,amsmath,amssymb,pra,aps,nofootinbib,superscriptaddress,floatfix]{revtex4-1}

\usepackage[T1]{fontenc}
\usepackage[utf8]{inputenc}
\usepackage{amsmath, amssymb}
\usepackage{bbold}
\usepackage{bm}
\usepackage{tipa}
\usepackage{subfigure}
\usepackage{blindtext}
\usepackage[dvipsnames]{xcolor}
\definecolor{dcyan}{RGB}{0,100,100}
\usepackage{graphicx}
\usepackage{tabularx}
\usepackage{textgreek}
\usepackage{xfrac}
\usepackage{units}
\usepackage{comment}
\usepackage{float}

\usepackage[sectionbib]{bibunits}
\defaultbibliographystyle{naturemag}
\defaultbibliography{references}

\usepackage{hyperref}
\usepackage{xcolor}
\definecolor{green_cust}{RGB}{0,154,85}
\definecolor{red_cust}{RGB}{173,49,54}
\definecolor{blue_cust}{RGB}{0,103,148}
\hypersetup{colorlinks=true, linkcolor=red_cust, citecolor=green_cust, filecolor=blue_cust, urlcolor=blue_cust}

\makeatletter 
    
\renewcommand\onecolumngrid{
\do@columngrid{one}{\@ne}%
\def\set@footnotewidth{\onecolumngrid}
\def\footnoterule{\kern-6pt\hrule width 1.5in\kern6pt}%
}

\renewcommand\twocolumngrid{
        \def\footnoterule{
        \dimen@\skip\footins\divide\dimen@\thr@@
        \kern-\dimen@\hrule width.5in\kern\dimen@}
        \do@columngrid{mlt}{\tw@}
}%

\makeatother    

\usepackage{soul}

\newcommand{\avg}[1]{\langle{#1}\rangle}

\newcommand{\Figref}[1]{Fig.~\hyperref[#1]{\ref{#1}}}

\begin{document}
\title{Cavity QED in a High NA Resonator}

\author{Danial Shadmany$^*$}
\affiliation{Department of Physics, Stanford University, Stanford, CA}
\author{Aishwarya Kumar$^*$}
\affiliation{Department of Physics, Stanford University, Stanford, CA}
\affiliation{Department of Physics and Astronomy, Stony Brook University, Stony Brook, NY}
\author{Anna Soper}
\affiliation{Department of Applied Physics, Stanford University, Stanford, CA}
\author{Lukas Palm}
\affiliation{Department of Physics, The University of Chicago and the James Franck Institute, , Chicago, IL}
\author{Chuan Yin}
\affiliation{Department of Physics, The University of Chicago and the James Franck Institute, , Chicago, IL}
\author{Henry Ando}
\affiliation{Department of Physics, The University of Chicago and the James Franck Institute, , Chicago, IL}
\author{Bowen Li}
\affiliation{Department of Physics, Stanford University, Stanford, CA}
\author{Lavanya Taneja}
\affiliation{Department of Physics, The University of Chicago and the James Franck Institute, , Chicago, IL}
\affiliation{Department of Physics, Stanford University, Stanford, CA}
\author{Matt Jaffe}
\affiliation{Department of Physics, The University of Chicago and the James Franck Institute, , Chicago, IL}
\affiliation{Department of Physics, Montana State University, Bozeman, MT}
\author{David I Schuster}
\affiliation{Department of Applied Physics, Stanford University, Stanford, CA}
\author{Jon Simon}
\affiliation{Department of Physics, Stanford University, Stanford, CA}
\affiliation{Department of Applied Physics, Stanford University, Stanford, CA}
\date{\today}

\begin{bibunit}


\begin{abstract}
From fundamental studies of light-matter interaction to applications in quantum networking \& sensing, cavity quantum electrodynamics (QED) provides a platform-crossing toolbox to control interactions between atoms and photons. The coherence of such interactions is determined by the product of the single-pass atomic absorption and the number of photon round-trips. Reducing the cavity loss has enabled resonators supporting nearly 1-million optical roundtrips at the expense of severely limited optical material choices and increased alignment sensitivity. The single-pass absorption probability can be increased through the use of near-concentric, fiber or nanophotonic cavities, which reduce the mode waists at the expense of constrained optical access and exposure to surface fields. Here we present a new high numerical-aperture, lens-based resonator that pushes the single-atom-single-photon absorption probability per round trip close to its fundamental limit by reducing the mode size at the atom below a micron while keeping the atom mm-to-cm away from all optics. This resonator provides strong light-matter coupling in a cavity where the light circulates only $\sim\hspace{-2pt}10$ times. We load a single $^{87}$Rb atom into such a cavity, observe strong coupling, demonstrate cavity-enhanced atom detection with imaging fidelity of $99.55(6)\%$ and survival probability of $99.89(4)\%$ in $130$~$\mu$s, and leverage this new platform for a time-resolved exploration of cavity cooling. The resonator's loss-resilience paves the way to coupling of atoms to nonlinear \& adaptive optical elements and provides a minimally invasive route to readout of defect centers. Introduction of intra-cavity imaging systems will enable the creation of cavity arrays compatible with Rydberg atom array computing technologies, vastly expanding the applicability of the cavity QED toolbox.
\end{abstract}
\maketitle

\section{Introduction}
\label{sec:intro}
Cavity quantum electrodynamics, the study of light-matter interaction through coupling of emitters to the field of an optical resonator, enables coherent information exchange between material and photonic degrees of freedom, with wide-ranging applications from qubit state detection~\cite{teper2006resonator,bochmann2010lossless,dhordjevic2021entanglement,deist2022mid}, to sensing~\cite{leroux2010implementation,cox2016deterministic,hosten2016measurement, pedrozo2020entanglement}, and networking~\cite{ritter2012elementary,reiserer2015cavity,kimble2008quantum}. Such tools apply not only to coupling of light to laser-cooled atoms and ions~\cite{birnbaum2005photon,steiner2013single,sterk2012photon, keller2004continuous}, but also transmon qubits, rare earth ionic dopants~\cite{raha2020optical,kindem2020control}, color centers~\cite{calusine2014silicon, evans2018photon,zhang2018strongly}, quantum dots~\cite{englund2007controlling, press2007photon}, and other optically active emitters.

The fundamental figure of merit that controls the coherence of light-matter interactions is the cooperativity, given by $C=\frac{4g^2}{\kappa\Gamma}$, where $g$ is the coherent information exchange rate, $\kappa$ is the decay rate of the cavity field and $\Gamma$ is the decay rate of the material excitation~\cite{tanji2011interaction}. For applications in quantum information science, the number of coherent information exchanges is approximately $\sqrt{C}$ (SI~\ref{SI:cavitytransfer}), and the infidelity of a cavity mediated gate is $\epsilon\approx\frac{2\pi}{\sqrt{C}}$(\cite{sorensen2003measurement,benito2019optimized} and SI~\ref{SI:cavitytransfer}). Cavities can also be harnessed for Purcell enhanced atomic state detection, where the ratio of atomic emission into the cavity vs free space is $C$~\cite{tanji2011interaction}; equivalently, the Purcell enhanced collection solid angle is $4\pi\times C$.

For a closed optical transition such as those available in alkali atoms, the cooperativity of a macroscopic resonator may be expressed in terms of the resonator geometry $C=\frac{12}{\pi^2}\frac{F}{2\pi}\frac{\lambda^2}{w^2}$ (\cite{tanji2011interaction} and SI~\ref{SI:CooperativityExpressions}), where $F$ is the resonator finesse defined so that $F/(2\pi)$ is the mean number of times the light passes the atom within the cavity, $w$ is the optical mode size at the atom, and $\lambda$ is the wavelength of the optical transition. At fixed $\lambda$, this expression suggests two routes to strong coupling ($C>1$)-- large finesse or small mode waist.

Finesses above $10^5$ have been achieved using ion-beam sputtered dielectric coatings on super polished~\cite{hood2001characterization} or laser ablated~\cite{hunger2010fiber} substrates, and recently finesses above $10^6$ have been achieved using reactive ion etching~\cite{jin2022micro} to reduce  surface roughness. Further gains would necessitate breakthroughs in atom-scale surface polishing and part-per-million-level coating absorption.

The decades-long effort to reduce the mode size $w$ has spanned numerous approaches: in two-mirror cavities, the near-confocal geometry~\cite{siegman1986lasers} with length $L$ equal to the mirror radius of curvature $R$ is a compromise between resonator alignment sensitivity and optical access that results in a mode-waist of $w=\sqrt{\frac{R\lambda}{2\pi}}$. Efforts to reduce the mode waist in two-mirror cavities beyond this limit have focused on either (a) small mirror separation $L\ll R$ (near-planar resonators), providing a substantial reduction in mode waist~\cite{birnbaum2005photon,hunger2010fiber} at the expense of reduced optical access and sensitivity to E-fields from proximal surfaces; or (b) near-concentric resonator geometry $L\approx 2R$, providing a few-fold waist reduction compared to the confocal geometry at the expense of increased alignment sensitivity~\cite{haase2006detecting,periwal2021programmable,deist2022mid}.

\begin{figure}[ht!] 
    \includegraphics[width=\columnwidth]{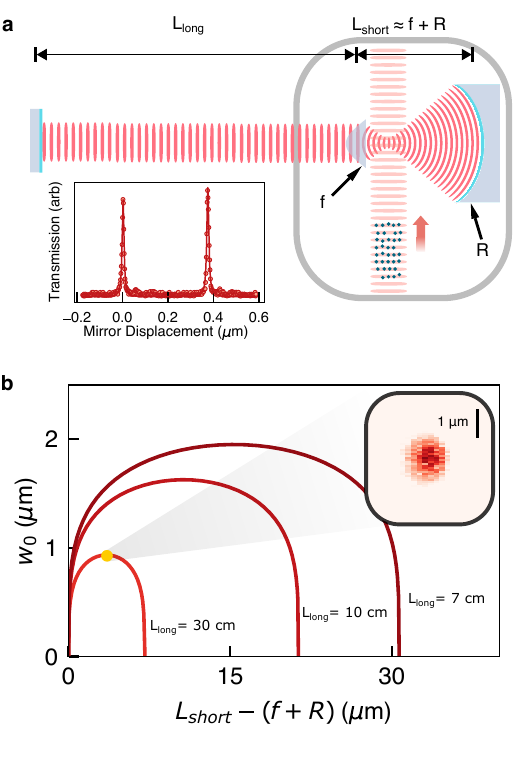}
    \caption{
        \textbf{Resonator Design}. The heart of the apparatus, shown in \textbf{(a)}, is a cavity consisting of a spherical mirror (radius $R$) that focuses light down to a sub-micron spot, a high-NA aspheric lens (focal length $f$) that re-collimates it, and finally a long propagation (length $L_{long}$). This long propagation distance necessitates that the vacuum chamber window (gray) reside inside of the resonator. An optical conveyor belt vertically transports $^{87}$Rb atoms to the resonator from a magneto-optical trap. \textbf{(inset)} A cavity transmission spectrum reveals a (fitted) finesse $F=40(2)$; for a 480 MHz free spectral range (FSR), the linewidth is thus $\kappa=2\pi\times 12.4(7)$~MHz. \textbf{(b)} The paraxially computed mode waist (w$_{0}$) is plotted vs the distance between the aspheric lens and spherical mirror, quantified as the deviation from overlapping lens and mirror foci. Introduction of the asphere within the cavity allows for the exploration of resonators that are asymmetric, consisting of two propagation arms of unequal lengths. For $L_{long} \gg L_{short}$, the width of the stability region ($f^2/L_{long}$) compresses, leading in the extreme case ($L_{long} \approx 30$~cm) to a sub-micron waist that persists even in the middle of the stable range, rather than exclusively near the (highly-sensitive) edges as is common in macroscopic two mirror cavities. \textbf{(inset)} A direct measurement of the mode waist (via scanning a nano-pore across the mode, see SI~\ref{SI:ModeWaistMeasurement}) bounds the mode size at a maximum $1/e^2$ radius of $1000$~nm.
    }

	\label{fig:ScalingFig}
\end{figure}

\begin{figure*}[ht] 
	\centering
 	\includegraphics[width=183 mm]{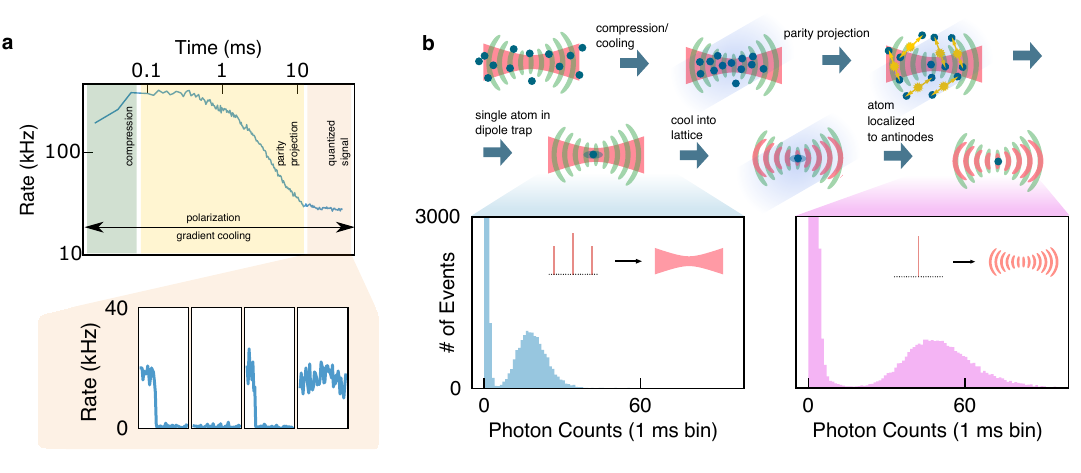}
	\caption{\textbf{Single Atom Loading} \textbf{(a)} Early time dynamics of the Rb cloud emission into the cavity. First (dark green), the signal increases as atoms are cooled into the cavity-enhanced 785 nm dipole trap. After reaching a maximum (yellow) the atoms at the center begin to parity project, leading to loss. \textbf{(inset)} Finally (orange) the signal settles to quantized levels, indicating the presence of 0 or 1 atoms. \textbf{(b)} Full schematic of single-atom loading and localization schemes. First, the atoms are released from the transport lattice  and cooled and compressed into the intracavity-dipole trap (red) by the 3D molasses beams (dark blue), where the intracavity-dipole trap is created by the addition of off-resonant sidebands exactly 10 FSRs away (see SI~\ref{SI:CavityDipoleTrap}). Near the small waist, the atoms in the dipole trap undergo light-assisted collisions, parity projecting until only 1 atom is left. In light blue, we show the histogram of scattering statistics with a 1 ms bins for a single atom in a dipole trap. This corresponds to a scattering rate of $\approx{20}$~kHz. The single atom in the dipole trap samples the nodes and anti-nodes readout cavity lattice (light green) equally, cutting the effective cooperativity by a factor of two. To improve on this, we alter the profile of the trapping light from a dipole trap to a lattice by removing the far sidebands. We take care to choose a 785 nm FSR which is phase-matched with the 780 nm readout lattice. Cooling is performed during this process, leading to a single atom which is localized to a single cavity anti-node. We plot a histogram of statistics for the localized single atom with a 1 ms bin size (pink) and find a scattering rate of $\approx{50}$~kHz. }
	\label{fig:AtomDet}
\end{figure*} 

Achieving yet-smaller mode waists requires exploring more sophisticated resonator geometries. For example, by employing an additional long propagation arm and/or a pair of \emph{convex} mirrors, the bowtie resonator allows for mode-waists down to a limit imposed by astigmatism induced by off-axis incidence on the focusing mirrors~\cite{ningyuan2016observation,clark2020observation}. Such resonators have demonstrated waists down to $\sim\!7$~$\mu$m~\cite{chen2022high}, the limit thus far achieved for high-finesse macroscopic Fabry-P\'erot cavities. Wavelength-scale resolutions have been observed in a multi-pass imaging system for biological samples~\cite{juffmann2016multi}, but never in a macroscopic device compatible with atomic cavity QED.

In this work we present, for the first time, a macroscopic resonator with a $\lambda$-scale mode waist. This enables us to enter the strong coupling regime of cavity QED in a resonator with finesse below $50$. In Section~\ref{sec:theplatform} we introduce this first-of-its-kind resonator geometry and describe its construction and stability. In Section~\ref{sec:coupling} we describe the process of loading/detecting a single atom and use it to characterize the resonator with a measurement of the atom-cavity coupling. In Section~\ref{sec:atomdetection} we demonstrate fast atom detection and in Section~\ref{sec:outlook} we conclude by describing opportunities to leverage this new approach for emerging quantum science and technology.

\section{Small Waist Lens Cavity}
\label{sec:theplatform}

The principal innovation in our resonator design is the integration of a high numerical aperture (NA) aspheric lens (focal length $f$) within a two mirror Fabry-Perot cavity. This lens divides the resonator into a short arm (length $L_{short}$) between the lens and a spherical mirror (radius of curvature $R$), and a long arm (length $L_{long}$) between the lens and a planar mirror (see Fig.~\ref{fig:ScalingFig}a). When $L_{long},R\gg f$, a small waist appears in the short arm near the center of curvature of the spherical mirror and focus of the aspheric lens. The size of this waist, $w_0\approx\sqrt{\frac{f}{L_{long}}}\times\sqrt{\frac{f\lambda}{2\pi}}$, is very similar to that of a confocal cavity, but with additional demagnification induced by the long arm. 


There are several important features of note: (i) if the long arm is \emph{long} enough, sub-micron waists are possible at the center of the stability diagram, limited by clipping and aberration due to finite lens NA (see SI~\ref{SI:LightCollectionComparison}. (ii) using an asphere allows us to minimize aberrations while maintaining a high numerical aperture (NA), unlike a bowtie resonator with spherical mirrors where astigmatism limits the achievable waist~\cite{ningyuan2016observation} (iii) with a round trip Gouy phase of $\pi/2$, the transverse mode structure of the cavity is neither that of a concentric or a confocal resonator, suppressing mode mixing due to lower order aberrations~\cite{jaffe2021aberrated}. For our parameters ($L_{long}= 30$~cm, $f= 1.45$~mm), we expect a waist of  $930$~nm  (see Fig.~\ref{fig:ScalingFig}b). See SI~\ref{SI:StabilityDiagrams}

In our implementation of the resonator geometry described above, we mount the asphere and the curved mirror in an ultra-high vacuum load-lock system (see SI Fig.~\ref{fig:flexure} and~\cite{yin2023cavity}). The mount is a custom designed 3-axis piezo-driven and mechanically multiplied flexure-stage, providing precise control of the relative position of the asphere and mirror, crucial to aligning the cavity in the presence of drifts (see SI~\ref{SI:Mechdesign}). We place the in-coupling (R=98$\%$) flat end mirror of the cavity outside vacuum and use it primarily to in-couple our probe and trapping light. We insert a pellicle right after the flat mirror to pick-off a controllable (angle-dependent) fraction of the light circulating in the cavity to either measure cavity transmission or atomic fluorescence. Furthermore, we employ an electro-optic modulator between the pellicle and vacuum window for high bandwidth cavity locking. We directly verify a micron scale waist in an out-of-vacuum test setup by using a gold coated film containing a $200$~nm diameter aperture (see SI~\ref{SI:ModeWaistMeasurement}).

\section{Atom-Cavity Coupling}
\label{sec:coupling}
To conclusively demonstrate high cooperativity and the utility of the platform in atom-based quantum information protocols, we trap a single atom at the waist of the resonator and use it to probe the cavity. Our experiments begin with a cloud of laser cooled $^{87}$Rb atoms from a magneto optical trap (MOT) which is transported to the small cavity waist location in a one-dimensional optical conveyor belt (see Fig.~\ref{fig:ScalingFig}a). Additional polarization gradient cooling (PGC) at the cavity waist location loads the atoms into an intra-cavity dipole trap at $785$~nm with a peak depth of $U_0/k_B = 2.0(5)$~mK (measured as half the trap-induced shift of the atom-cavity resonance condition).

\begin{figure}[ht] 
	\centering
 	\includegraphics[width=\columnwidth]{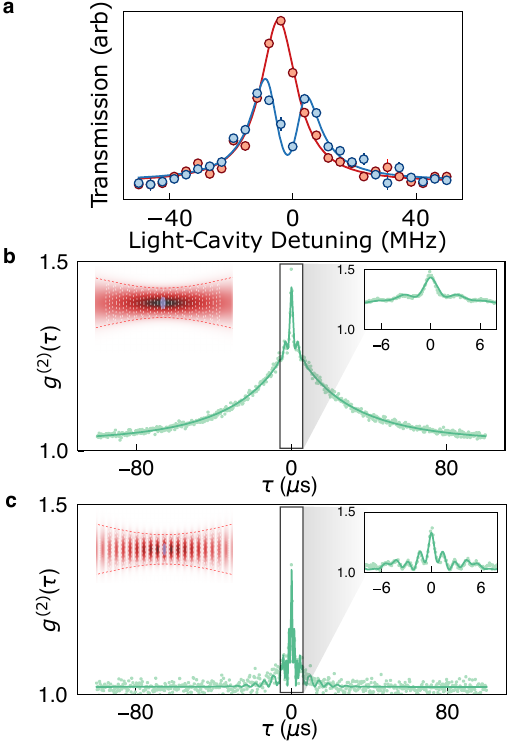}
	\caption{
	\textbf{Trapped Atom Characterization.} Once an atom is detected, we probe its cavity coupling: \textbf{(a)} Cavity transmission vs probe laser frequency, with/without atom in orange/purple; the mode splitting apparent in the presence of the atom indicates strong light-matter coupling with $C=1.6(2)$. \textbf{(b)} Two-time correlator $g^{(2)}(\tau)$ of cavity-fluorescence from a dipole-trapped atom vs time difference $\tau$. The fast ($\sim 3~\mu$s) oscillations \textbf{(inset)} come from radial oscillations away from the cavity axis where coupling is strongest, rapidly damped by PGC. The slow decay over $\sim 50~\mu$s comes from over-damping the weak axial dipole trap. A fit (solid, see SI~\ref{SI:g2}), indicates axial \& radial temperatures differ by a factor $\sim 100$ ($T_{z}\approx 5~\mu$K, $T_{x,y}\approx 600~\mu$K), while axial \& radial PGC-damping rates differ by a factor $\sim 10$ ($\gamma_{x,y}=500$~ms$^{-1}$, $\gamma_z=5000$~ms$^{-1}$). \textbf{(c)} $g^{(2)}(\tau)$ of lattice trapped atom; absence of slow decay indicates stronger axial trapping of the lattice. The fit (solid, see SI~\ref{SI:g2}), reveals equal axial \& radial temperatures $=200~\mu$K, with axial \& radial damping rates differing by a factor of four ($\gamma_{x,y}=112$~ms$^{-1}$, $\gamma_z=400$~ms$^{-1}$). \textbf{b} \& \textbf{c}, \textbf{insets} depict trapping potential in red, light-matter coupling in white, and extent of atomic motion in purple.
	}
	\label{fig:VRS}
\end{figure}

It is essential that atoms within the cavity be able to collide to ensure that the PGC process drives parity projection~\cite{ueberholz2002cold}. As such, the atoms are first loaded into a dipole trap rather than a lattice. We wash out the cavity standing wave by applying sidebands to the $785$~nm cavity lattice laser, to excite cavity modes 10 free spectral ranges away, creating a dipole trap at the small waist location (see SI \ref{SI:CavityDipoleTrap}). Once a single atom is loaded into the dipole trap, we remove the sidebands while cooling over $2$~ms, reloading the atom into a single cavity lattice well. The trapping wavelength must thus be chosen to align the $780$~nm and $785$~nm cavity standing-waves at the location of the small cavity waist (see SI~\ref{SI:LatticeAlignment}). To probe the system we monitor the PGC-induced fluorescence scattered into the cavity, picked off by the pellicle and directed into a pair of single photon counting modules (SPCMs).

We first study the atom-cavity interaction with outcoupling $T=4\%$ well below the internal cavity round trip loss $L_{rt}=11.7\%$ to achieve a near-maximal finesse of $\mathcal{F}=40.0(1)$, and hence a predicted single-atom cooperativity $C_{max}=6$. Fig.~\ref{fig:AtomDet}a shows averaged fluorescence as an atomic ensemble is cooled into the intra-cavity dipole trap. An initial compression leads to an increase in the fluorescence, followed by a sharp decrease during parity-projection, after which the signal settles to quantized levels (Fig.~\ref{fig:AtomDet}a). This quantized signal arises from discrete atomic occupancy of the cavity trap, dominated by either 0 or 1 atom loaded~\cite{schlosser2001sub}. Fig.~\ref{fig:AtomDet}b shows the histogram of collected fluorescence photon numbers from dipole trapped atoms, with a clear separation between the background and a single atom loaded. After transferring to an intracavity lattice with depth $800(80)~\mu$K, the signal is enhanced 2-fold because the atom is better localized to an antinode of the $780$~nm cavity field (see SI~\ref{SI:LatticeAlignment}). 

Armed with the ability to condition measurements on the presence of an atom, we extract the single-atom cavity coupling strength by measuring the vacuum Rabi splitting. In Fig.~\ref{fig:VRS}a, we plot the cavity transmission as we scan a $780$ nm probe laser across the cavity resonance, both with (purple) and without (orange) an atom. A single-atom vacuum Rabi splitting is clearly visible with a fitted coupling strength $g=2\pi\times 5.6(3)$~MHz. Combined with the atomic linewidth $\Gamma=2\pi\times 6.065$~MHz and our measured $\kappa=2\pi\times13.3(1)$~MHz, this yields a cooperativity, $C=1.6$(2), placing us in the regime where emission into the cavity is more likely than into free space. This cooperativity is about a factor of $\sim 4$ lower than the maximum achievable due to a combination of cavity birefringence ($\approx 2\times$ reduction) and atomic motion ($\approx 2\times$ reduction) (see SI~\ref{SI:avgcoupling}). We also leverage our ability to perform atom-conditioned measurements to extract the second order correlation function, $g^{(2)}(\tau)$ of the atomic fluorescence, taken both in the dipole trap (Fig.~\ref{fig:VRS}b) and lattice (Fig.~\ref{fig:VRS}c) (see Supplement ~\ref{SI:g2dataconditions} for parameters). Both plots exhibit $\mu$s-scale oscillations indicative of radial trapping that is damped out over a few periods by the optical-molasses (see SI~\ref{SI:g2} for derivation of the solid fits); this oscillation reflects the fact that atoms are more likely to scatter at times when they are near cavity axis (large $g_2(\tau=0)$), and less likely to scatter a half trap period later when the atom has oscillated away from the cavity axis (small $g_2(\tau=T_{trap}/4)$). Data from the dipole-trapped atom exhibits an additional slow decay indicative of over-damping of the weakly-trapped axial motion. By contrast, the lattice-trapped atom is strongly confined in the axial direction, and so exhibits an additional, even faster (axial) oscillation in the $g_2$ instead place of over-damped decay apparent for the dipole-trapped atom.

\section{Optimized Atom Detection}
\label{sec:atomdetection}
The low outcoupling probability employed for the aforementioned experiments is not optimal for atomic state detection in the presence of fixed internal cavity losses. The optimum outcoupling is a tradeoff: increased outcoupling improves the fraction of cavity photons leak out of the cavity before they are lost to scattering/absorption of cavity optics, while reduced outcoupling improves the cooperativity and thus the probability that the atom scatters light into the cavity in the first place. As we show in SI~\ref{SI:outcoupling}, the optimal operating point occurs when the outcoupling probability is $L_{int}\sqrt{1+C_{max}}$, where $C_{max}$ is the cooperativity with no outcoupling and $L_{int}$ is the internal cavity loss per round trip. To further demonstrate the flexibility of our platform, we enter this optimal regime by tuning the angle of the intra-cavity pellicle to outcouple more light (optimized at 20 $\%$, with an estimated new cooperativity of $\approx{0.8}$).

\begin{figure}[ht!] 
	\centering
 	\includegraphics[width=\columnwidth]{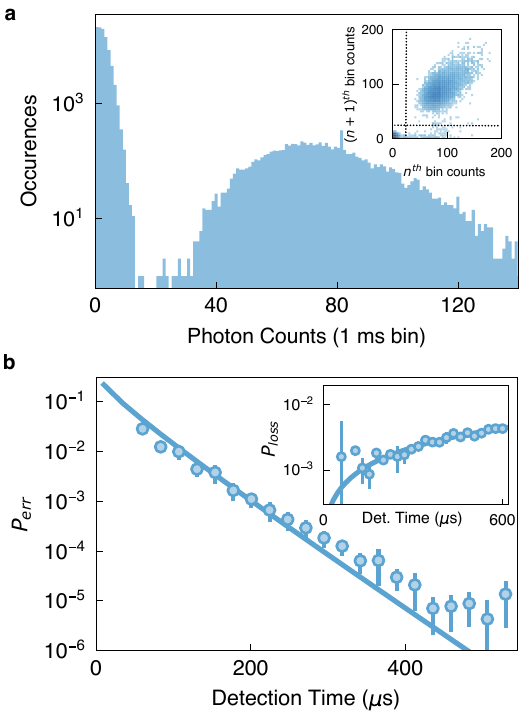}
	\caption{\textbf{Characterizing Atom Detection Fidelity.}
	\textbf{(a)} Histogram of cavity fluorescence photons collected in $1~$ms bins from a single lattice-trapped atom at larger $T=20\%$ cavity outcoupling. The measured scattering rate is $\sim 75$~kHz, $1.7\times$  higher than that measured in the same configuration at $T=4\%$ outcoupling. See Supplement~\ref{SI:outcoupling} for derivation of relationship between total cavity collection rate and outcoupling. \textbf{(inset)} Correlations between subsequent measurements of a single atom: upper-right quartile reflects persistence of an atom through adjacent measurements; lower-right quartile reflects an atom lost between the measurements; lower-left quantile reflects measurements both with no atom; upper-left quantile reflects appearance of an atom between measurements, indicating imperfect dispersal of the transported cloud. The elliptical shape of the data in the upper-right quadrant indicates correlations in scattering rate between subsequent measurements, likely due to site to site variation in cavity coupling strength. \textbf{(b)} The model-independent fidelity estimates are plotted vs measurement time, showing a fidelity of $99.55(6)\%$ reached at a measurement time of $130~\mu$s. The \textbf{(inset)} depicts the corresponding estimated loss probability vs the imaging time and with a model (solid curve) for a survival rate of  99.89(4)$\%$ and a$112(8)$~ms lifetime (see SI \ref{SI:lifetie}).
	}
	\label{fig:Fidelities}
\end{figure} 

Figure~\ref{fig:Fidelities}a shows atom detection for optimized outcoupling. Compared to the low-outcoupling configuration, we observe an increase in the detected photon rate by a factor of $\sim$1.7 (Fig.~\ref{fig:Fidelities}a), consistent with expectations (see SI~\ref{SI:outcoupling}). A systematic, model-independent analysis of atom detection (see Ref.~\cite{norcia2018microscopic} and SI~\ref{SI:DetEfficelModelFree}) reveals that, within 130~$\mu$s, we are able to detect an atom with fidelity of $99.55(6)\%$ (see Fig.~\ref{fig:Fidelities}b), while maintaining an atom survival probability of 99.89(4) $\%$ (see Fig.~\ref{fig:Fidelities}b inset), consistent with the atom lifetime in the presence of the molasses light. These numbers are limited by the internal losses of our resonator, without which we could increase the resonator finesse, and thereby the Purcell enhancement factor of the resonator, while maintaining efficient resonator outcoupling. Reducing these losses, through a combination of better anti-reflection coatings and lower surface roughness, will immediately reduce the readout time.

\section{Outlook}
\label{sec:outlook}
In this paper we have introduced an approach to cavity QED that leverages high NA, low-finesse cavities to enter the strong coupling regime. We load a single atom into such a cavity, with a finesse $F=40$, harnessing cavity enhancement to create first a dipole trap and then a lattice, which efficiently confines the atom. We validate that we are in the strong coupling regime via an atom-conditioned vacuum Rabi splitting measurement, proving the presence of a single atom via a quantized signal. Temporal correlations of the light scattered by the atom into the cavity enable precise measurements of both atom temperature and PGC damping coefficients. When we optimize the cavity outcoupling to maximize total atomic light scattering, we are able to detect a single atom with a fidelity of 99.55(6)\% in a time of $130~\mu$s, with a survival rate of $99.9\%$.

The parameters of our platform are far from optimized: a non-birefringent cavity will provide $2\times$ the collection efficiency, while lower atom temperature will provide an additional factor of $1.8$; free-space photon counters provide an extra factor $2$ vs their fiber-coupled counterparts, for a projected detection time of $20~\mu$s. Reduced intracavity loss will yield further improvements in collection efficiency and cooperativity, enabling non-destructive atom detection in less than $\sim 10~\mu$s. See SI~\ref{SI:LossBudget}. The minimal finesse requirements mean that it will be possible to strongly couple the atom to more sophisticated intracavity optics: nonlinear crystals will enable in-situ wavelength conversion for integration with telecom infrastructure~\cite{chen2005intracavity}; electro-optics will provide rapid tunability~\cite{hudson2005mode}; adaptive optics will enable yet-higher NA operation~\cite{benea2021electro}; and nanophotonics in the end-mirror will enable direct integration with waveguide devices. 

Equally exciting is the possibility of integrating these cavities with solid-state emitters that otherwise require nanoscale cavities to achieve large cooperativity due to inherent material-induced scattering/absorption: coupling to rare-earth ions and silicon vacancies, with their nearly-closed transitions, would greatly increase the flexibility of those platforms. Finally, in combination with microlens arrays to stabilize an array of waists, it should be possible to extend the small-waist resonator technique demonstrated in this paper to \emph{arrays of resonators spaced by a few microns}, for immediate integration with Rydberg atom arrays~\cite{endres2016atom,barredo2016atom}. Crucially, because atoms can be mm-to-cm distances from the nearest optic, the sensitivity of Rydberg atoms to surface potentials is mitigated in these cavities compared to their nanophotonic counterparts. In short, this work heralds a new era of optical cavity QED where strong light-matter coupling is widely available, rapidly extensible, and compatible with a much broader array of experimental platforms and technologies.

\section{Acknowledgments}
This work was supported by AFOSR grant FA9550-22-1-0279, ARO grant W911NF-23-1-0053, AFOSR MURI Grant FA9550-19-1-0399, and AFOSR DURIP FA9550-19-1-0140. D.S. and H.A. acknowledge support from the NSF GRFP. A.S. acknowledges support from the Hertz Foundation and the DoD NDSEG Fellowship. We acknowledge Mark Stone for assistance in construction of the load-lock vacuum system, and Adam Shaw for discussions of model-free estimates of atom detection fidelity.

\section{Author Contributions}
The experiments were designed by A.K., D.S., A.S., M.J., L.P., D.I.S., and J.S. The apparatus was built by C.Y., H.A., M.J., D.S., A.S., A.K., L.T., and J.S. The collection of data was handled by A.K., D.S. and A.S. L.P. and B.L. performed the waist measurements in a test setup. All authors analyzed the data and contributed to the manuscript.

$^*$ These authors contributed equally.

\section{Competing Interests}
The authors declare no competing financial or non-financial interests.

\clearpage


\putbib
\end{bibunit}


\subsection{Data Availability}
The experimental data presented in this manuscript are available from the corresponding author upon request, due to the proprietary file formats employed in the data collection process.
\subsection{Code Availability}
The source code for simulations throughout are available from the corresponding author upon request. 
\subsection{Additional Information}
Correspondence and requests for materials should be addressed to J.S. (jonsimon@stanford.edu). Supplementary information is available for this paper.

\clearpage
\newpage





\begin{bibunit}
\onecolumngrid
\newpage
\section*{Supplementary Information}
\appendix
\renewcommand{\appendixname}{Supplement}
\renewcommand{\theequation}{S\arabic{equation}}
\renewcommand{\thefigure}{S\arabic{figure}}
\renewcommand{\figurename}{Supplemental Information Fig.}
\renewcommand{\tablename}{Table}
\setcounter{figure}{0}
\setcounter{table}{0}
\numberwithin{equation}{section}

\section{Derivation of Out-scattering Efficiency}
\label{SI:outcoupling}
For a cavity with cooperativity $C_{max}$ in the absence of any outcoupling (just intracavity losses), the question we want to address here is how much outcoupling to \emph{add} to maximize the fraction of photons scattered by the atom that leak out of the cavity: if we outcouple too little, more of the photons are lost to intracavity losses; if we outcouple too much we reduce the cooperativity and more photons are lost to free-space scattering.

Suppose we outcouple a factor $\beta$ more photons than are lost intracavity. This provides a cavity outcoupling fraction $\frac{\beta}{1+\beta}$ and a cavity cooperativity $C=\frac{C_{max}}{1+\beta}$ The fraction of photons scattered by the atom that leak out of the cavity for detection is then $\chi=\frac{\beta}{1+\beta}\times\frac{C_{max}/(1+\beta)}{1+C_{max}/(1+\beta)}$. Maximizing this over $\beta$ yields $\beta=\sqrt{1+C_{max}}$, and $\chi=1-2\frac{\sqrt{1+C_{max}}-1}{C_{max}}$.

In the limit of large maximum cooperativity $C_{max}\gg 1$, we then find that $C\approx\sqrt{C_{max}}$ and $\chi\approx 1-\frac{2}{\sqrt{C_{max}}}$; on the other hand, in the limit of small cooperativity $C_{max}\ll 1$, we find that $C\approx\frac{C_{max}}{2}$ and $\chi\approx \frac{C_{max}}{4}$.

That is, in the limit of large $C_{max}$ we should outcouple enough to reduce the cooperativity to the square root of its maximum value, resulting in outcoupling which is $\approx\sqrt{C_{max}}$ times larger than the intracavity losses. By contrast small $C_{max}$ we should set our outcoupling equal to our intracavity losses, halving the cooperativity.

These results are interesting because common knowledge (see SI~\ref{SI:cavitytransfer}) was that coherent transfer of photons between atoms has an infidelity (for $C\gg 1$) of approximately $\frac{2\pi}{\sqrt{C}}$, whereas scattering light into a cavity (thought of as an incoherent process) has an infidelity of approximately $\frac{1}{C}$; rather than attributing this difference to coherent vs incoherent scattering, it is now clear that the difference arises from \emph{neglecting cavity outcoupling efficiency} in the latter case. This is valuable to know, because otherwise it would have appeared more efficient to pitch and catch quantum information between atoms in separate cavities rather than coherently transferring between atoms in the same cavity - it is now apparent that these two processes have the same (unfavorable) scaling with the maximum cooperativity of the resonator.
\begin{figure}[ht]
	\centering
 	\includegraphics[width=180 mm]{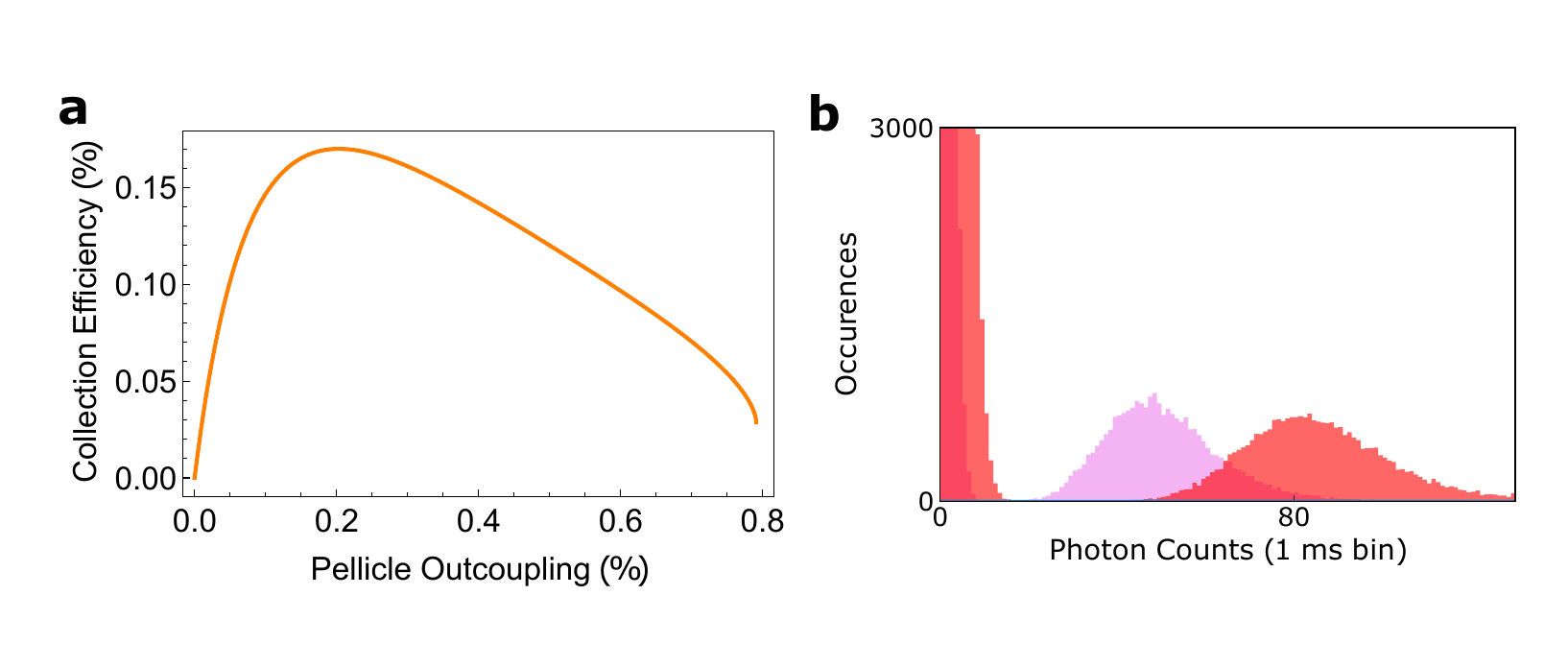}
	\caption{\textbf{Cavity Collection Efficiency.}
	\textbf{(a)} Photon collection efficiency as function of outcoupling fraction for measured cooperativity \textbf{(b)} histogram showing improvement between increase in outcoupling from 4$\%$ (pink) to 20$\%$ (red)
	}
	\label{fig:Cav}
\end{figure} 

\section{Cavity-Mediated Coupling Between Ensembles}
\label{SI:cavitytransfer}
This section follows the approach outlined in Jon Simon's thesis~\cite{simon2010cavity}. Suppose we have two atoms ($a$ \& $b$) in the same cavity and want to coherently couple a hyperfine excitation (between states $g$ \& $f$) from one to the other via a cavity-stimulated Raman transition. The Hamiltonian is:
\begin{equation}
    H=g \left[c^\dagger (\sigma^{fe}_a+\sigma^{fe}_b)+c (\sigma^{ef}_a+\sigma^{ef}_b)\right]+\Omega (\sigma^{eg}_a+\sigma^{eg}_b+\sigma^{ge}_a+\sigma^{ge}_b)+(\delta_c+i\frac{\kappa}{2}) c^\dagger c+(\Delta+i\frac{\Gamma}{2}) (\sigma^{ee}_a+\sigma^{ee}_b)
\end{equation}

Here the cavity is detuned from Raman resonance by an amount $\delta_c$, the Raman transition is detuned from the excited state by an amount $\Delta$, $g$ is the single photon/single atom light-matter coupling strength on the $f\leftrightarrow e$ transition, $\Omega$ is the Rabi frequency for the laser driving the $g\leftrightarrow e$ transition, $\Gamma$ \& $\kappa$ are the excited atom and cavity decay rates, $c\dagger$/$c$ are the field creation and annihilation operators, and $\sigma^{pq}_l$ is the transition operator from state $q$ to state $p$ in atom $l$.

There are, of course many pulse protocols to convert $|fg;0\rangle \leftrightarrow |gf;0\rangle$. In the presence of dominant cavity and excited atom loss, they all have the same optimal performance up to factors of order unity. For simplicity we will skip counter-intuitive driving and all other adiabatic approaches (though they are covered in Simon's thesis), and consider only a 4-photon Rabi process.

In that case we can adiabatically eliminate the $e$-states, resulting in an effective 2-photon Hamiltonian (neglecting second order shifts that can be compensated with cavity \& laser detunings):
\begin{equation}
    H^{eff}_2=\frac{g\Omega}{\Delta} \left[c^\dagger (\sigma^{fg}_a+\sigma^{fg}_b)+c (\sigma^{gf}_a+\sigma^{gf}_b)\right]+i\frac{\Omega^2}{\Delta^2}\Gamma(\sigma^{ff}_a+\sigma^{ff}_b)+(\delta_c+i\kappa/2+i \frac{g^2}{\Delta^2}\Gamma) c^\dagger c
\end{equation}

Further eliminating the cavity mode results in an effective 4-photon Hamiltonian (neglecting fourth order shifts that can be tuned out):
\begin{equation}
    H^{eff}_4=\frac{g^2\Omega^2}{\Delta^2\delta_c} \left[\sigma^{fg}_a\sigma^{gf}_b+\sigma^{gf}_a\sigma^{fg}_b\right]+i\left[\frac{\Omega^2}{\Delta^2}\Gamma+\frac{g^2\Omega^2}{\Delta^2\delta_c^2}(\kappa/2+\frac{g^2}{\Delta^2}\Gamma)\right](\sigma^{ff}_a+\sigma^{ff}_b)
\end{equation}

A full flip-flop of an excitation between atoms $a$ and $b$ thus requires a time $\tau=\pi/\left(\frac{g^2\Omega^2}{\Delta^2\delta_c}\right)$, resulting in a total decoherence induced loss $\epsilon=\tau\times\left[\frac{\Omega^2}{\Delta^2}\Gamma+\frac{g^2\Omega^2}{\Delta^2\delta_c^2}(\kappa/2+\frac{g^2}{\Delta^2}\Gamma)\right]$. In the limit $\Delta\rightarrow\infty$, this becomes $\epsilon=\pi\frac{2\Gamma\delta_c^2+g^2\kappa}{2g^2\delta_c}$. This loss is minimized for $\delta_c=\frac{g\sqrt{\kappa}}{\sqrt{2\Gamma}}=\kappa\sqrt{\frac{C}{8}}$, with a value $\epsilon=\frac{2\pi}{\sqrt{2C}}$.

Because any two-atom gate can be implemented by~\cite{pellizzari1995decoherence} mapping two physically separate qubits into one four-level atom via the cavity field,  performing internal four-level operations on the atom, and then mapping back to separate physical qubits via the cavity field, $\epsilon$ sets a lower limit on the infidelity of such a gate.

Similarly, the number of information exchanges between atoms, or between atom and cavity field, is $1/\epsilon\propto\sqrt{C}$.

Finally, a bit of interpretation: if we go to too large a detuning from the cavity, the 4-photon Rabi oscillation slows down too much and we are dominated by excited-state loss; if we go to too small a detuning from the cavity, the 4-photon Rabi oscillation speeds up, but not as much as the cavity leakage, and we are dominated by that. The sweet spot is in the middle. The excited state detuning, as long as it is large enough, does not contribute at all. Indeed one can show that even on resonance with excited state a counter-intuitive pulse sequence gives the same performance up to a numerical factor of order unity (see ref.~\cite{simon2010cavity}).

\section{Cavity Optics Setup}
\label{SI:ExpSetup}

\begin{figure*}[t] 
	\centering
 	\includegraphics[width=0.7\textwidth]{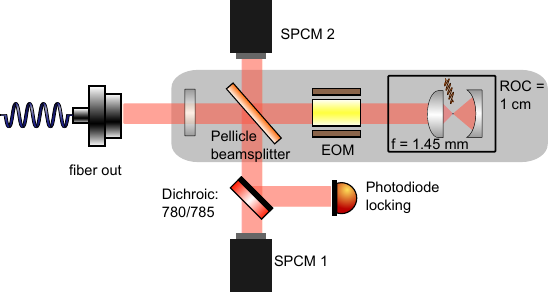}
	\caption{\textbf{Detailed Cavity Setup.} The resonator with all in-cavity optics is highlighted in gray. In addition to the aspheric lens, spherical mirror, and planar mirror, we also depict the pellicle used for outcoupling, and the intra-cavity electro-optic modulator that we use to lock the resonator with $\sim 100$~kHz of bandwidth. Light outcoupled from both sides of the pellicle is coupled to single photon counting modules (SPCMs); a dichroic on one side of the outcoupler picks off the 785~nm locking light.}
	\label{fig:supSetup}
\end{figure*}

See Figure \ref{fig:supSetup} for a full schematic, including intra-cavity optics, photo-detectors, and in-coupling paths. The setup includes an EOM and a pellicle beam splitter inserted into the long propagation arm of the cavity. These are used for high-bandwidth cavity stabilization and tunable outcoupling respectively. During operation, we tune the angle of the pellicle to switch between 4$\%$ and 20$\% $ pick-off, depending on if we are optimizing for atom-light coupling (as for a VRS) or overall photon collection efficiency (as for the readout histograms). The pellicle reflects light out on both sides, requiring the use of two SPCMs to maximize collection. Using the same pellicle, we also pick off a fraction the 785 nm trap light for a low-bandwidth, slow locking path, from which we generate an error signal and feedback to the piezo and EOM.

\section{Intracavity Dipole Trap from Multichromatic Driving}
\label{SI:CavityDipoleTrap}
We form our intra-cavity trap by driving the cavity with a far detuned $785$~nm laser. Driving a single longitudinal mode of the cavity results in a one-dimensional optical lattice with a small waist, which after parity projection, typically leads to single atoms trapped at multiple sites of the lattice. Deterministic trapping of only one atom coupled to the cavity mode requires an optical tweezer potential -- a dipole trap with a small waist. To create this dipole trap, we drive multiple longitudinal modes of the cavity by phase modulation of the trapping light. Since different longitudinal modes are phase shifted (in space) with respect to each other, choosing appropriate phase modulation frequency and strength can result in significant suppression of the sinusoidal intensity variation across a finite region~\cite{cox2016spatially}. For our cavity, we require this region to be centred at $d=1$~cm from the curved mirror -- the location of the small waist. Phase modulation with an EOM at $n$ times the free spectral range (FSR) results in a trapping potential given by (to the first order):

\begin{equation}
    U = U_0 \left(J_{-1}(\beta)^2 \sin^2(kz-\frac{nd}{L} \pi)+J_{0}(\beta)^2 \sin^2(kz)+J_{1}(\beta)^2 \sin^2(kz+\frac{nd}{L} \pi) \right)
\end{equation}

where $U_0$ is the depth of the initial optical lattice, $k$ is the wave-vector of the carrier, $\beta$ is the modulation depth, $L\approx30$~cm is the length of our cavity, and $J_{\alpha}$ is the Bessel function of first kind. We note the following identity :

\begin{equation}
\sin^2(kz-\frac{10}{30} \pi)+ \sin^2(kz)+\sin^2(kz+\frac{10}{30} \pi) = \frac{3}{2}
\end{equation}

\begin{figure*}[ht] 
	\centering
 	\includegraphics[width=0.95\textwidth]{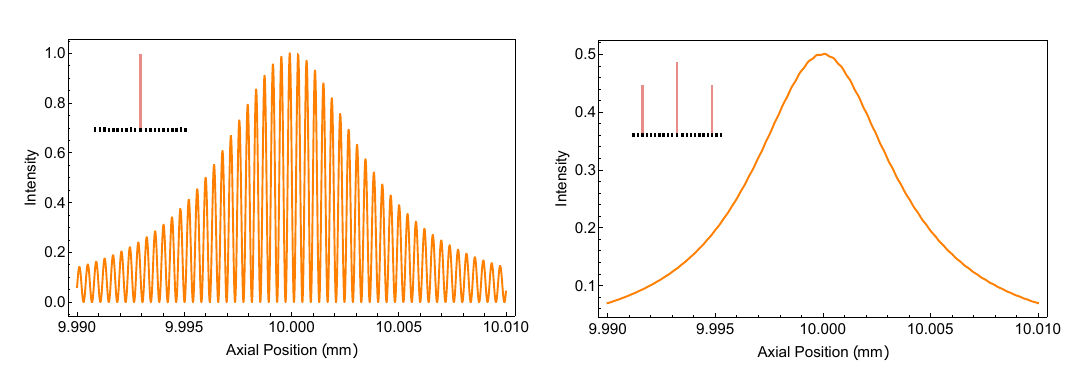}
	\caption{
		\textbf{left} the intensity profile of a single longitudinal mode of the cavity, which takes the form of standing wave alternating with nodes and anti-nodes \textbf{right} a smooth intensity profile is required to trap a single atom using light assisted collisions at the small waist. To realize this in the cavity, we drive sidebands 10 FSRs from the carrier, tuning the relative power in the peaks until the multiple longitudinal modes interfere to create a dipole trap. 
	}
	\label{fig:IntracavDipoleTrap}
\end{figure*}

For our cavity $L/d\approx30$, thus driving the $n=10$ sidebands at equal intensity as the carrier (i.e. $J_{-1}(\beta)^2 = J_0(\beta)^2 = J_{1}(\beta)^2$), results in a cancellation of the sinusoidal potential at the cavity waist resulting in a dipole trap with depth $U_0/2$ (see Fig.~\ref{fig:IntracavDipoleTrap}). While this seems to imply that such a cancellation is only achievable with a fine tuning of the length and the waist location, numerical calculations suggest that it is possible for any location by changing the modulation depth. Note that the effect of higher order sidebands can be similarly compensated for by slightly changing the modulation depth.

\section{Aligning the 780 nm cavity mode and 785 nm intra-cavity lattice}
\label{SI:LatticeAlignment}

Getting the strongest possible coupling between the atom and the cavity requires that the atom be trapped at the $780$~nm cavity mode anti-node. We achieve this  by using an intra-cavity lattice for trapping the atom, and choosing the frequency of the lattice laser such that the anti-nodes of lattice standing wave and the cavity mode are approximately aligned at $z=d$, where $d\approx1$~cm is the distance of the mode anti-node closest to the cavity waist, from the curved mirror \footnote{Note that we can't guarantee that there is an anti-node at the small waist location. At worst, the anti-node is $\lambda/4$ away from the waist, which only changes the coupling by $0.5\%$}.

To see how we find the correct lattice frequency, note that the lattice standing wave is given by the function $\sin^2(\frac{n d}{L})$, where $n$ is the order of the mode at which the lattice laser is resonant with the cavity, i.e. $L=n\lambda/2$, $\lambda\approx785$~nm. Moving this standing wave one period requires changing the phase the frequency by $\Delta n\times$FSR, where $\Delta n=\frac{L}{d}$. Our measured FSR gives $L=31.2$~cm, so $L/d$ is not an integer and therefore perfect alignment is not generally possible. But this guarantees that moving the frequencies by 16 FSRs would either pass through either a minimum or maximum of coupling. 

One of the ways we keep track of this coupling is by measuring the VRS. Another, faster way to keep track of this coupling, is to trap the particle in the dipole trap with a small residual standing wave, which still leads to parity projection, but slightly increases the probability of finding an atom at the residual lattice intensity anti-nodes. This leads to change in scattering rate in the dipole trap as the lattice laser is moved through the different longitudinal modes. We have used both methods at different times.

We then use a binary search like procedure to find the minimum of the coupling, since that is a clearer signal than a maximum and then change the frequency by 16 FSRs to find the maximum. Finally, we note that since smallest step that we can change the phase of the standing wave is $\frac{d}{L}\pi$, theoretically the worst possible phase offset (if we perform our alignment procedure perfectly) we can have is $\frac{d}{2L}\pi=0.016 \pi$, which corresponds to a misalignment of $\approx 6$~nm.

\section{Single Atom Lifetime}
\label{SI:lifetie}

\begin{figure*}[ht] 
	\centering
 	\includegraphics[width=0.5\textwidth]{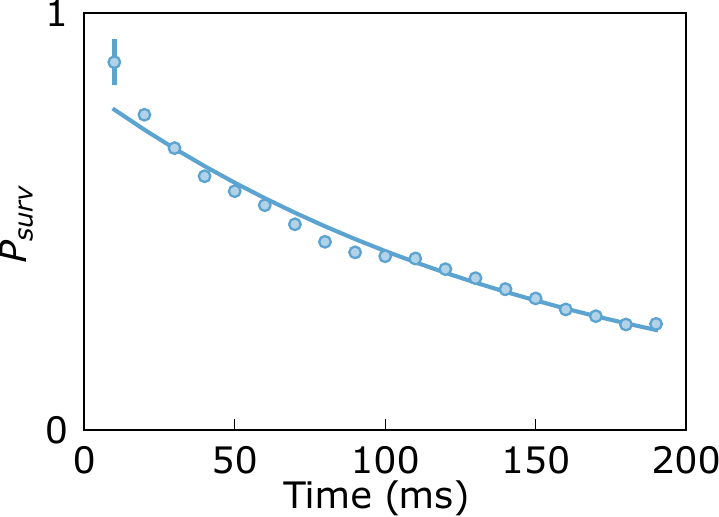}
	\caption{Survival probability of single atoms in the trap vs hold time, characterized by average fluorescence level over 700 shots, with lifetime fitted to 146(6)~ms.}
	\label{fig:lifetime}
\end{figure*}

To characterize the lifetime of our trapped single atoms, we observe the atom while under molasses cooling light for 200 ms and plot average fluorescence level vs hold time. An exponential fit then yields a lifetime of 146(6)~ms. See Figure ]\ref{fig:lifetime}.

The atom lifetime without the molasses light (in the dark) fits to a lifetime of 50~ms. The nearly threefold increase in lifetime in the presence of molasses light is likely indicative of intensity-noise-induced parametric heating in the cavity dipole trap/lattice.

\section{Experimental Conditions for $g_2(\tau)$ dataset}
\label{SI:g2dataconditions}
The $g_2(\tau)$ dataset was collected at high outcoupling ($T=20\%$ per round-trip) to maximize the data rate, using a pair of single photon counters (SPCMs) to minimize the impact of detector afterpulsing~\cite{mckeever2004trapped}, and time-tagged using custom firmware running on a RedPitaya built on top of the Zynq Time-to-Digital package (\cite{adamivc2019fast} and \url{https://github.com/madamic/zynq_tdc}).


\section{Cavity Stability}
\label{SI:StabilityDiagrams}

\begin{figure*}[ht] 
	\centering
 	\includegraphics[width=0.95\textwidth]{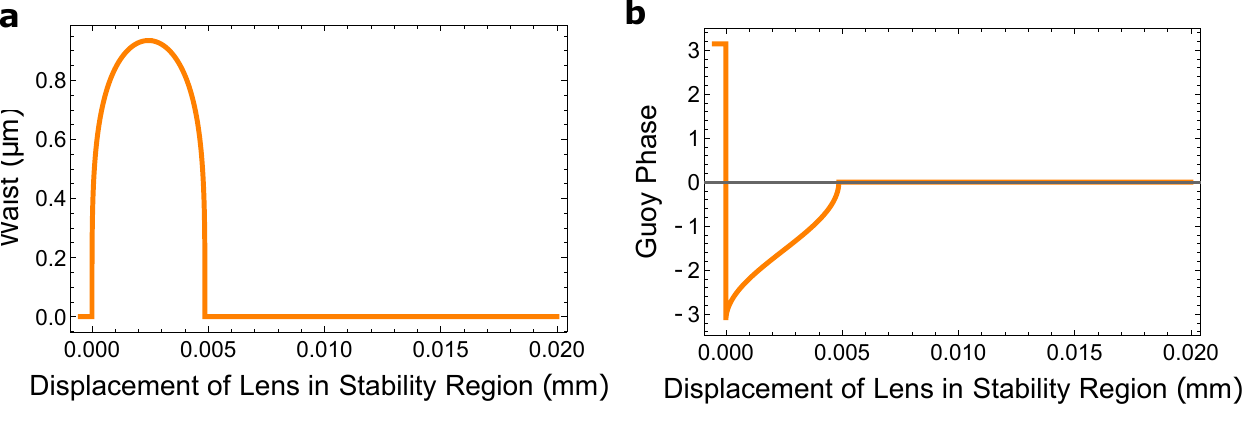}
	\caption{
		\textbf{left} the small waist cavity stability diagram with a long arm $L_{long} = 30$~cm. A waist of 930 nm is achieved at the center of the stability diagram \textbf{right} the Guoy phase of the cavity across the stability range, sweeping from $\pi$ to 0.
	}
	\label{fig:stabilitysup}
\end{figure*}

We calculate the waist size and Guoy phase of the small-waist resonator in the usual way, by taking the eigenvalues and vectors of the round trip ABCD matrix. Expressed as (1,q), the eigenvector provides the complex beam parameter q from which the waist size at a given cavity length can be extracted. The stability condition $\frac{f^2}{L_{long}-f}>L_{short}-f-R>0$ is extracted by setting the norm of the eigenvalue equal to 1. The argument of this round trip eigenvalue provides the Guoy phase, or transverse mode splitting. We show plots of the resonator waist size and Guoy phase at the operating point of our cavity, where we set the long arm $L_{long} = 30$~cm and the the focal length $f=1.45$~mm for the C140TMD-B asphere. The only varying parameter then is the length of the short arm (distance between C140 and curved mirror). We plot the waist size against the perturbation in this short arm length from $f$ (with $f$ chosen as the edge of the stability diagram and the ``design'' distance of a normal lens system). See Figure \ref{fig:stabilitysup}.

It is worth noting that the small waist in \textbf{a} occurs at the center of the stability diagram. This is the salient feature of our new resonator geometry, indicating that a sub-micron mode waist can be achieved at the location of the atom without operating at the edge of the stability range, as is required in concentric cavities. It is also worth noting at this point that the Guoy phase for this resonator does not follow the usual bounds for a two mirror cavity. There, the Guoy phase and consequently the transverse mode splitting runs from 0 to $2 \pi$ across the edges of the stability region. In the small waist geometry sketched here, the phase range is halved, running only from $\pi$ on one end to 0. This is due to the asymmetry of the cavity. At the operation point of our cavity - the center of the stability diagram - the Guoy phase of the resonator is $\frac{\pi}{2}$. This has the favorable property that neither the first or second order modes are overlapped with the fundamental mode of the cavity, circumventing the mode-mixing issues of the confocal and concentric designs.


\section{Error and Uncertainty Calculations}
\label{SI:errorbars}

Error propagation for quoted parameters is carried out using Python's uncertainties package, with error for fitted parameters provided by the covariance matrix. For uncertainty bars, two methods are used: (1) chi-squared, fit based for the lifetime data, and (2) bootstrapping for the survival rate and fidelity data. For bootstrapping, we split the data into three sets, calculate relevant quantities individually, then calculate the average and standard deviation.

\section{$g^{(2)}(\tau)$ Calculation}
\label{SI:g2}

There are three relevant timescales in this system : (i) the fastest one set by the Rabi frequency in PGC, detuning, atomic decay rates and the cavity lifetime, (ii) the time scale set by the scattering rate in the PGC, and (iii) the slowest timescale, set by the motion in the trap. While it might seem that the fastest, quantum mechanical timescale would be most interesting, it has been thoroughly explored in the context of resonance fluorescence. With time-resolved explorations of laser-cooling in mind, we are interested in the slowest of these timescales. Here the dynamics of the atom are entirely described by the classical Langevin equation of a harmonic oscillator. The 1D Langevin equation is of the form :

\begin{equation}
\label{eqn:langevin}
 \frac{d^2 x(t)}{dt^2} + \gamma \frac{dx(t)}{dt} + \omega^2 x(t) = \eta(t)
\end{equation}

Here \( \eta(t) \) is the random force (noise term) that obeys $\avg{\eta(t)\eta(t')}= \frac{2\gamma k_B T}{m}\delta(t-t')$, \( \gamma \) is the damping rate, $m$ is the mass, \( \omega \) is the oscillator angular frequency, and $T$ is the temperature.

Our program for analyzing these dynamics is as follows:

\begin{enumerate}
\item Treat the atom as moving in a (separable) intra-cavity potential $U(x,y,z)=\frac{U_0}{1+\left(\frac{z}{z_r}\right)^2}\times \exp\left(-\frac{2(x^2+y^2)}{w_0^2(1+(z/z_r)^2)}\right)\approx U_0\times\left(1-(\frac{z}{z_r})^2\right)\times\left(1-2\frac{x^2+y^2}{w_0^2}\right)\approx U_0\times\left(1-(\frac{z}{z_r})^2-2\frac{x^2+y^2}{w_0^2}\right)$, where $z_r\equiv\frac{\pi w_0^2}{\lambda}\approx 3$~$\mu$m is the Rayleigh range of the cavity mode, $w_0\approx 930$~nm is the mode waist, and $\lambda\approx 780$~nm is the wavelength of the trapping light.
\item Treat the atomic scattering rate as separable in the Cartesian atom coordinates $I_{sc}\approx I_{pk}\times\cos^2 {k z} \times e^{-2(x^2+y^2)/w_0^2}$. Note that where the \emph{trap} varies in $z$ on a length scale of $z_r$, the scattering rate varies in $z$ on a much shorter length-scale of $k^{-1}=\frac{\lambda}{2\pi}$ -- by contrast, the radial trapping and radial scattering rates both vary on the length scale of $w_0$ - it is for this reason that we claim that the variations in fluorescence rate in the dipole trap arise predominantly from axial motion\footnote{At least insofar as axial and radial temperatures are equal}.
\item Generate the equations of motion for each of the spatial degrees of freedom of the atom, including velocity dependent damping and Langevin noise term arising from the cooling beams (itself normalized based upon the axis-dependent temperature).
\item For the radial direction, solve these equations for the two-time correlators of each degree of freedom of the atom.
For the axial direction, the fast $\cos^2(k z)$ variation of the scattering rate requires calculation of the the marginal probability distribution $P(z,t)$
\item Employ these two-time correlators and the marginal distribution of the atomic position, in conjunction with the knowledge that since the equations of motion are linear the motional degrees of freedom act as gaussian processes, to compute two-time correlators of the intensity. We will find that the two-time intensity correlator factorizes into independent correlators in $x$,$y$, and $z$.
\item Normalize the two-time correlator using the expected value of the intensity.
\end{enumerate}

We first note that since motion in the three directions is approximately independent, the variations of scattering rates due to motion in the three directions are uncorrelated. Writing scattered intensity as $I(t) = I_{pk}I_x(x(t))I_y(y(t))I_z(z(t))$, we find :

\begin{equation}
\begin{aligned}
g^{(2)}(\tau) &= \frac{\avg{I_x(x(\tau))I_y(y(\tau))I_z(z(\tau))I_x(x(0))I_y(y(0))I_z(z(0)}}{\avg{I_x(x(0))I_y(y(0))I_z(z(0)}^2} \\
&= \frac{\avg{I_x(x(\tau))I_x(x(0))}}{\avg{I_x(x(0))}^2}
\frac{\avg{I_y(y(\tau))I_y(y(0))}}{\avg{I_y(y(0))}^2}
\frac{\avg{I_z(z(\tau))I_z(z(0))}}{\avg{I_z(z(0))}^2}\\
&=g^{(2)}_x(\tau)g^{(2)}_y(\tau)g^{(2)}_z(\tau)
\end{aligned}
\end{equation}

\subsection{The radial part}
In one of the radial directions, say $x$, the scattered intensity varies as $I_x(x) = e^{-\frac{2x^2}{w_0^2}} \approx 1-\frac{2x^2}{w_0^2}$. This implies :

\begin{equation}
    g_x^{(2)}(\tau) = \frac{\biggl<\left(1-\frac{2x(\tau)^2}{w_0^2}\right)\left(1-\frac{2x(0)^2}{w_0^2}\right)\biggr>}{\biggl<1-\frac{2x(0)^2}{w_0^2}\biggr>^2}
\end{equation}

We assume a gaussian process and use Wick's theorem to simplify the fourth order correlator and obtain:

\begin{equation}
    g_x^{(2)}(\tau) = 1+\frac{8}{w_0^4}\frac{\avg{x(\tau)x(0)}^2}{\left(1-\frac{2\avg{x(0)^2}}{w_0^2}\right)^2}
\end{equation}

We further note that for a particle governed by equation~\ref{eqn:langevin}, the second order position correlator is well known and given by Ref~\cite{wang1945theory}:

\begin{equation}
    \avg{x(\tau)x(0)} = \frac{k_B T_{r}}{m \omega_r^2}e^{-\frac{\gamma_r}{2} \lvert\tau\rvert}\left(\cos(\omega' \tau)+\frac{\gamma_r}{2\omega'}\sin(\omega'\lvert\tau\rvert)\right),
\end{equation}

\noindent where $T_r$ is the radial temperature, $\omega_r$ is the radial trap angular frequency, $\gamma_r$ is the radial damping rate, and $\omega'=\sqrt{\omega_r^2-\gamma_r^2/4}$. This completes the calculation for $g^{(2)}_x(\tau)$. We assume that our system has cylindrical symmetry, so $g^{(2)}_y(\tau)= g^{(2)}_x(\tau)$.

\subsection{The axial part}

In the axial direction $z$, the variation in scattered intensity is due to the standing wave nature of the mode, i.e. $I_z(x)=\cos^2(kz)$, leading to:

\begin{equation}
    g^{(2)}_z(\tau)=\frac{\avg{\cos^2(kz(\tau))\cos^2(kz(0))}}{\avg{\cos^2(kz(0))}^2}
\end{equation}

\subsubsection{Intra-cavity lattice}
For an atom trapped in tashe intra-cavity lattice, we can follow the same procedure as the radial calculation i.e. Taylor expand the cosine, and apply Wick's theorem, yielding :

\begin{equation}
    g_x^{(2)}(\tau) = 1+2 k^4\frac{\avg{z(\tau)z(0)}^2}{\left(1-k^2 \avg{z(0)^2}\right)^2}
\end{equation}

\subsubsection{Dipole trap}
This is not possible for an atom in the dipole trap, which has a length scale given by $z_r$, the Rayleigh range. An atom could possibly sample even multiple longitudinal mode anti-nodes while staying in the trap.

Here we instead use the fact that the trap frequency in dipole trap is much lower, and assume that the motion of the particle is overdamped. This allows us to simplify the Fokker-Planck equation and solve for the marginal probability distribution, $P(z,t;z_0)$, the probability of finding a particle at position $z$ at time $t$ given that the particle was at position $z_0$ at time $t=0$. The Fokker-Planck equation simplifies to~\cite{wang1945theory}:

\begin{equation}
    \frac{\partial P(z, t)}{\partial t} = \frac{\partial}{\partial z} \left( \frac{\omega_{ax}^2}{\gamma_{ax}} z P(z, t) \right) + \frac{k_B T_{ax}}{m \gamma_{ax}} \frac{\partial^2 P(z, t)}{\partial z^2}
\end{equation}

Here $T_{ax}$ is the axial temperature, $\omega_{ax}$ is the axial trap angular frequency, $\gamma_{ax}$ is the radial damping rate. Such a simplification and marginalization of the full distribution including velocity is possible because the particle attains terminal velocity at every position, giving $\avg{v}=\frac{\omega_{ax}^2}{\gamma_{ax}}z$. This allows the inertial term in the Langevin equation to be neglected. For the initial condition $P(z,0)=\delta(z-z_0)$, the solution to this equation is a gaussian given by:

\begin{equation}
    P(z,t;z_0)=\frac{1}{\sigma(t)\sqrt{2\pi}}\exp\left(-\frac{(z-z_0 \alpha(t))^2}{2\sigma(t)^2}\right),
\end{equation}

\noindent with $\alpha(t)=e^{-\frac{\omega_{ax}^2}{\gamma_{ax}}t}$, $\sigma(t) = \sigma_0 (1-\alpha(t)^2)$ and $\sigma_0=\frac{k_B T_{ax}}{m \omega_{ax}^2}$. $P(z,t;z_0)$ is thus the conditional probability of finding the atom at position $z$ at time $t$, given that it was at position $z_0$ at time $t=0$, exactly the distribution required to calculate two-time correlators. Furthermore, $P_{\infty}(z) = P(z,\infty;z_0)$ gives the probability of finding an atom at position $z$ in steady state.

For the denominator of $g^{(2)}_z(\tau)$ we have:
\begin{equation}
\begin{aligned}
\avg{\cos^2(kz)}^2 &= \left(\int_{-\infty}^{\infty}dz P_{\infty}(z)\cos^2(kz)\right)^2 \\
&= \frac{1}{2}\left(1+e^{-2k^2\sigma_0^2}\right)^2
\end{aligned}
\end{equation}

For the numerator we get:
\begin{equation}
\begin{aligned}
\avg{\cos^2(kz(\tau)\cos^2(kz_0)} &= \int_{-\infty}^{\infty}\left(dz_0 P_{\infty}(z_0)\cos^2(kz_0)\int_{-\infty}^{\infty}dz P(z,t;z_0) \cos^2(k z(\tau))\right) \\
&= \frac{1}{4} \left(e^{-4 k^2 \sigma_0^2} \cosh \left(4 \alpha(\tau)  k^2 \sigma_0^2\right)+2 e^{-2 k^2 \sigma_0^2}+1\right)
\end{aligned}
\end{equation}

Putting it all together and simplifying with a bit of algebra, we get for the dipole trap:
\begin{equation}
    g^{(2)}_z(\tau) = 1+\frac{2 \sinh ^2 \left(2 k^2 \sigma_0^2 e^{-\left( \frac{\omega_{ax}^2}{\gamma_{ax}}\lvert\tau\rvert \right)}\right)}{\left(e^{2 k^2 \sigma_0^2}+1\right)^2}
\end{equation}

\section{Effect of temperature on average atom-cavity coupling}
\label{SI:avgcoupling}
The geometric expression for the cooperativity mentioned in the main text is valid for a standing wave cavity with the atom perfectly localized to a cavity mode anti-node at the cavity waist. In reality, due to its finite temperature, the atom samples a range of different positions and therefore different couplings to the mode. This leads to an average observed cooperativity lower than the peak value. We note that we are neglecting the effect of the atom sampling different velocities, Doppler broadening the atomic transition. We expect this effect to be much smaller than the lowering of effective coupling, even at the highest expected temperatures.

The functional dependence of cooperativity on the position is exactly the same as that of scattered intensity discussed in SI~\ref{SI:g2}. In fact, if the average value of the cooperativity is reduced from the peak cooperativity ($C_{peak}$ by a factor f), i.e. $\avg{C} = f C_{peak}$, the factor f is given by exactly the kind of integrals we have already calculated for the denominator of $g^{(2)}$ in SI~\ref{SI:g2}. We are only interested in the cooperativity when the atom is trapped in the lattice. In this case the reduction factor is given by:

\begin{equation}
    f =    \biggl(1-\frac{2\avg{x^2}}{w_0^2}\biggr)\biggl(1-\frac{2\avg{y^2}}{w_0^2}\biggr)\left(1-k^2 \avg{z^2}\right)
\end{equation}

As before, $x$ and $y$ are the radial directions and $z$ is the axial direction,

\begin{equation}
\begin{aligned}
   \avg{x^2}&=\avg{y^2}=\frac{k_B T_{r}}{m \omega_r^2}\\
   \avg{z^2}&=\frac{k_B T_{ax}}{m \omega_{ax}^2}
\end{aligned}
\end{equation}
This gives,
\begin{equation}
    f=\biggl(1-\frac{k_B T_r}{2 U_0}\biggr)^2\left(1-\frac{k_B T_{ax}}{2 U_0}\right)
\end{equation}

Here $U_0$ is the depth of the trap. In this expression we have neglected the slight difference between $785$~nm trap profile and the $780$~nm mode profile and also assumed perfect alignment between both of them. For our measured values of $U_0/k_B\approx 800$~$\mu$K and $T_r\approx T_{ax}\approx 200$~$\mu$K, we get $f\approx 0.67$. Based on our waist, $w_0 = 930$~nm and finesse, $F=40$, $C_{peak}=5.6$. Accounting for another factor of 2 from lack of optical pumping and probing with a linear polarization, we predict $\avg{C}=1.87$, close to the measured value of $1.6$. This difference could potentially be explained by the effect of stochastic loading of the atom in different sites along the $z$ direction on the scale of a Rayleigh range, which would lead to additional variation in coupling, which we have neglected.

\section{Mode Waist Measurement}
\label{SI:ModeWaistMeasurement}
Approaching the diffraction limit $1/kw_0\rightarrow 1$, the smallest achievable cavity waist is influenced by several non-quadratic optical effects: non-paraxial propagation \& aberrations of the asphere~\cite{jaffe2021aberrated} and vector polarization effects~\cite{chaumet2006fully} that are small in free space are enhanced by numerous cavity round-trips. Rather than relying exclusively on modeling to estimate the beam-waist, we prefer to measure it directly.

A traditional knife edge measurement of the waist is not possible, because once the losses from the edge become comparable to the external and internal losses of the cavity (long before it actually approaches the waist), the finesse drops precipitously. In addition, the short ($z_r\sim 3~\mu$m) Rayleigh range of the sub-micron waist necessitates measurement with an object that is thin compared to $z_r$, impractical for a razor blade. Indeed, even SNOM-based mode-mapping techniques~\cite{ferri2020mapping} result in too much loss near the waist of a sub-micron waist resonator.

We therefore introduce a new way of measuring the mode waist of a cavity in situ using an end mirror with a small, sub-wavelength hole providing local outcoupling of the cavity field. The size of the hole is chosen such that the additional loss introduced is small compared to the cavity round trip loss and the finesse does not drop significantly for all measurement locations, while still providing enough outcoupling to detect the transmission. We use a gold coated SiN membrane (Norcada) with two holes of nanopores (of radius $100$, $200$, or $350$ nm) separated by $7~\mu$m. The gold coated membrane acts as the flat end mirror of the cavity which guarantees the focus to be right on the membrane because of the flat wavefront set by the boundary conditions. We carried out our test by using aspheric lens with $f=1.5$~mm (Thorlabs C140 TMD-B), identical to what we used in the main text.

The transmission through a sub-wavelength nanopore is approximated by the Bethe formula~\cite{Bethe1944} $T \propto (\lambda/r)^4$ for a circular hole with radius $r$ at wavelength $\lambda$ in a thin and transversely infinite conductor. In practice the finite thickness of the gold deposition as well as the existence of surface plasmons and plasmon resonances around the hole complicate the calculation of the hole transmission; because of the $r^{-4}$ dependence of transmission on hole size, the size must be carefully chosen to ensure that the transmission is smaller than the mirror transmission $T=1-98.5 \%=0.015$ to avoid spoiling the cavity finesse, but large enough to yield a detectable signal. Experimentally we found a diameter of $200$~nm gave a clear transmission signal while only reducing the cavity finesse by $15 \%$ from 52 to 42, whereas $350$~nm introduced excess loss destroying the cavity and $100$~nm yielded no transmission detectable above the noise floor. We expect that the impact of the $200$~nm pinhole's size on our waist measurement to be ignorable. The hole is then scanned by a homebuilt, mechanically multiplied 2D piezo stage transverse to the cavity axis to determine the profile of the cavity waist. The well-defined separation of the two holes is used as an absolute calibration of the scanning setup.

\begin{figure}[ht]
	\centering
 	\includegraphics[width=180 mm]{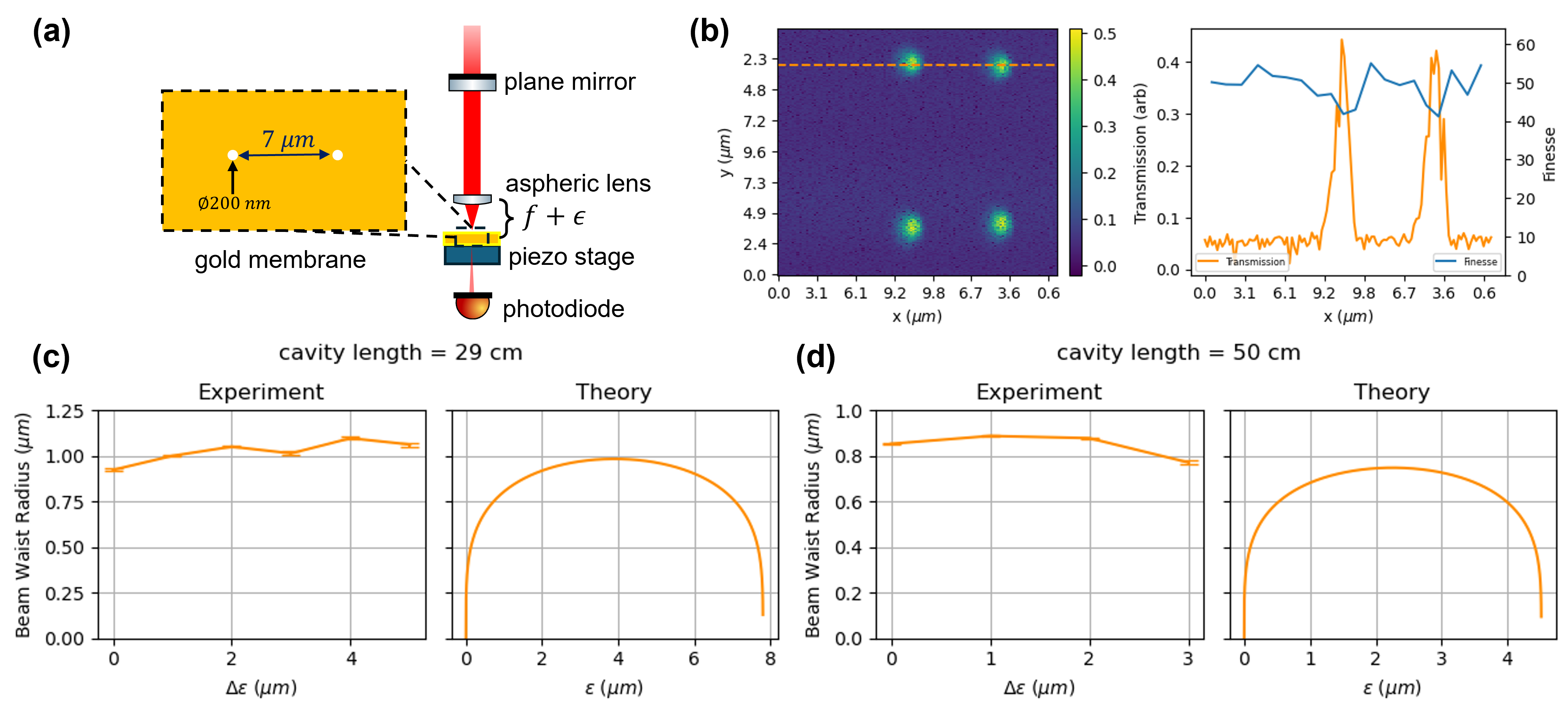}
	\caption{\textbf{Measurement on Mode Waist.}
	\textbf{(a)} Schematic of measurement setup. We built a similar cavity setup as in the main text and placed a gold membrane with pinhole with 200 nm diameter and pitch of $7~\mu $m on a piezo scanning stage behind the aspherical lens. The extra distance $\epsilon$ is tunable via the translation stage holding the asphere. The piezo stage scans the 2D transverse plane as the transmission signal is recorded by a photodiode. \textbf{(b)} 2D Plot of transmission. As cavity mode crosses the pinhole, the finesse drops only around 15\% from 52 to 42, which allows the transmission measurement to reflect the point spread function (PSF) without a substantial impact to finesse variation/impedance matching. Note that the duplication of pinhole pairs on the upper and lower regions is due to the triangle scanning waveform in the $y$ direction of piezo stage. \textbf{(c)-(d)} Comparison of experiment and paraxial model when cavity length is $29$~cm in (c) and $50$~cm in (d). Note that the $x$ axis of experiment plots $\Delta \epsilon$ refers to relative displacements in $\epsilon$ while the $x$ axis $\epsilon$ in theory plots refers to absolute displacements. We expect the absolute $\epsilon$ values in the experiment overlap with the central parts of theory $\epsilon$ values. The scale in x direction was calibrated by using $7~\mu$m pitch between the pinholes and we only took the gaussian fit in the $x$ direction as our data. The error bar is the standard deviation of the average of 4 PSF features in each transmission image.
	}
	\label{fig:MoveWaistMeasurement}
\end{figure}

In figure \ref{fig:MoveWaistMeasurement} the transmission through the nanopore is shown as a function of transverse position. The waist is determined by a gaussian fit to both holes.

We extract the waist size for several different positions in the stability diagram, as a function of mirror lens separation. We measure the finesse and amplitude by sweeping the laser over multiple FSRs at every transverse position. From the higher order modes in the transmission spectrum we infer the transverse mode spacing and the position in the stability diagram.

\section{Comparing Lens and Cavity Light Collection Rates}
\label{SI:LightCollectionComparison}
We would like to compare the collection efficiency of a high numerical aperture  lens (parameterized here by $\mathrm{NA}_{lens}$) to that of a standing wave cavity with waist size also limited by $\mathrm{NA}_{lens}$.

The first thing to note is that if all we care about is maximizing the cavity cooperativity and our loss comes entirely from clipping, it is \emph{always} better to reduce the cavity NA by increasing the mode waist to improve the finesse. The idea is that the finesse limit is set by clipping four per round-trip to $F_{cav}=\frac{2\pi}{4 e^{-2\mathrm{NA}_{lens}^2/\mathrm{NA}_{cav}^2}}$, where $\mathrm{NA}_{cav}=\frac{\lambda}{\pi w_0}$ and $w_0$ is the cavity mode waist. The cavity cooperativity is then: $C=\frac{24F}{\pi}\frac{1}{(k w)^2}=\frac{6F}{\pi}\mathrm{NA}_{cav}^2=3\mathrm{NA}_{cav}^2 e^{2\mathrm{NA}_{lens}^2/\mathrm{NA}_{cav}^2}$

Because the gaussian factor grows faster than the quadratic prefactor shrinks, we should go to low $\mathrm{NA}_{cav}$ (large cavity waist) to maximize the cooperativity!

What went wrong? Well, for starters we might \emph{want} a small mode waist for other reasons, like coupling to atoms in an array, or because achieving high cooperativity at small mode waist relaxes finesse requirements and speeds up readout. The upshot is that, if we do not expect to be able to get our finesse higher than a few thousand (for example due to material or linewidth constraints), choosing $\mathrm{NA}_{cav}\approx \frac{1}{2}\mathrm{NA}_{lens}$ is plenty, as it leads to a finesse of approximately 4600; increasing the cavity NA rapidly reduces the finesse -- choosing $\mathrm{NA}_{cav}\approx \frac{3}{4}\mathrm{NA}_{lens}$ reduces the finesse to 54.

With the choice $\mathrm{NA}_{cav}\approx \frac{1}{2}\mathrm{NA}_{lens}$ out of the way, we now assume we actually end up with a substantially \emph{lower} finesse limited by material constraints (glass absorption, imperfect AR coatings, etc...); in the absence of cavity outcoupling, the maximum achievable finesse is assumed to be $F_{max}$; choosing to an outcoupler with transmission $\beta$ times the internal cavity losses reduces the finesse to $\frac{F_{max}}{1+\beta}$, but outcouples a fraction $\frac{\beta}{1+\beta}$ of the light within the cavity.

As such, if, in the absence of the cavity, an atom scatters into $4\pi$ at a rate $\Gamma_{4\pi}$, the cavity will collect light and outcouple it at a rate $\Gamma_{cav}=\Gamma_{4\pi}\times \frac{6F_{max}}{\pi (1+\beta)}\times\mathrm{NA}_{cav}^2\times\frac{\beta}{1+\beta}$. If we optimize this rate over $\beta$, we find that we should choose $\beta=1$, halving the cavity finesse, resulting in 
$\Gamma_{cav}=\Gamma_{4\pi}\times \frac{3F_{max}}{2\pi}\times\mathrm{NA}_{cav}^2$. Noting that $\mathrm{NA}_{cav}\approx \frac{1}{2}\mathrm{NA}_{lens}$ we have: $\Gamma_{cav}=\Gamma_{4\pi}\times \frac{3F_{max}}{8\pi}\times\mathrm{NA}_{lens}^2$.

We should now compare this to the single-lens collection efficiency. A lens with $\mathrm{NA}_{lens}\ll 1$ subtends a solid angle $\Omega_{lens}\approx \pi \mathrm{NA}_{lens}^2$. This means that the lens collects light at a rate $\Gamma_{lens}=\Gamma_{4\pi}\times\frac{\Omega_{lens}}{4\pi}=\Gamma_{4\pi}\times \frac{\mathrm{NA}_{lens}^2}{4}$.

We can now compute the cavity collection efficiency enhancement $R=\frac{\Gamma_{cav}}{\Gamma_{lens}}=\frac{3F_{max}}{2\pi}\approx\frac{F_{max}}{2}$. That is, the total improvement in collection that the cavity provides is given by half of its finesse.

\emph{Note: We have neglected polarizations, assumed that cavity scattering does not induce extra heating, and worked for the lens in the low NA limit. These assumptions can be made most accurate with an optically pumped atom at a cavity anti-node, detected in cavity transmission.}

\section{Cavity Structure Mechanical Design}
\label{SI:Mechdesign}

\begin{figure}[ht]
	\centering
 	\includegraphics[width=180 mm]{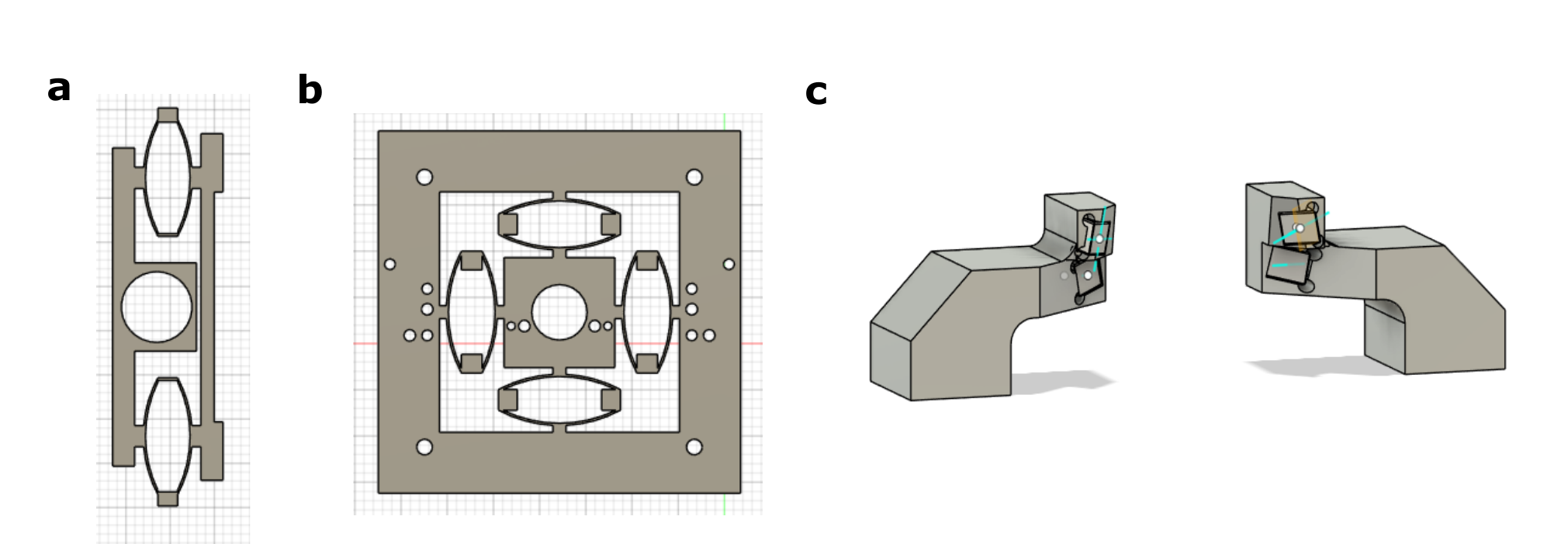}
	\caption{\textbf{Molasses mounts.}
	\textbf{(a)} The vertical flexure allows for control of the y-position of the curved mirror. Piezo-electric stacks slide into both holders depicted here. \textbf{(b)} 2D flexure allows for control of the x and z positions of the aspheric lens. Together with the vertical flexure, all degrees of freedom in the relative alignment between the asphere and curved mirror are accounted for. Piezo stacks slide into only half off the holders here. \textbf{(b)} Molasses mounts, with 3D angles set by proper alignment of beams to the small waist.
	}
	\label{fig:MolassesMounts}
\end{figure}

\begin{figure*}[b] 
	\centering
 	\includegraphics[width=0.45\textwidth]{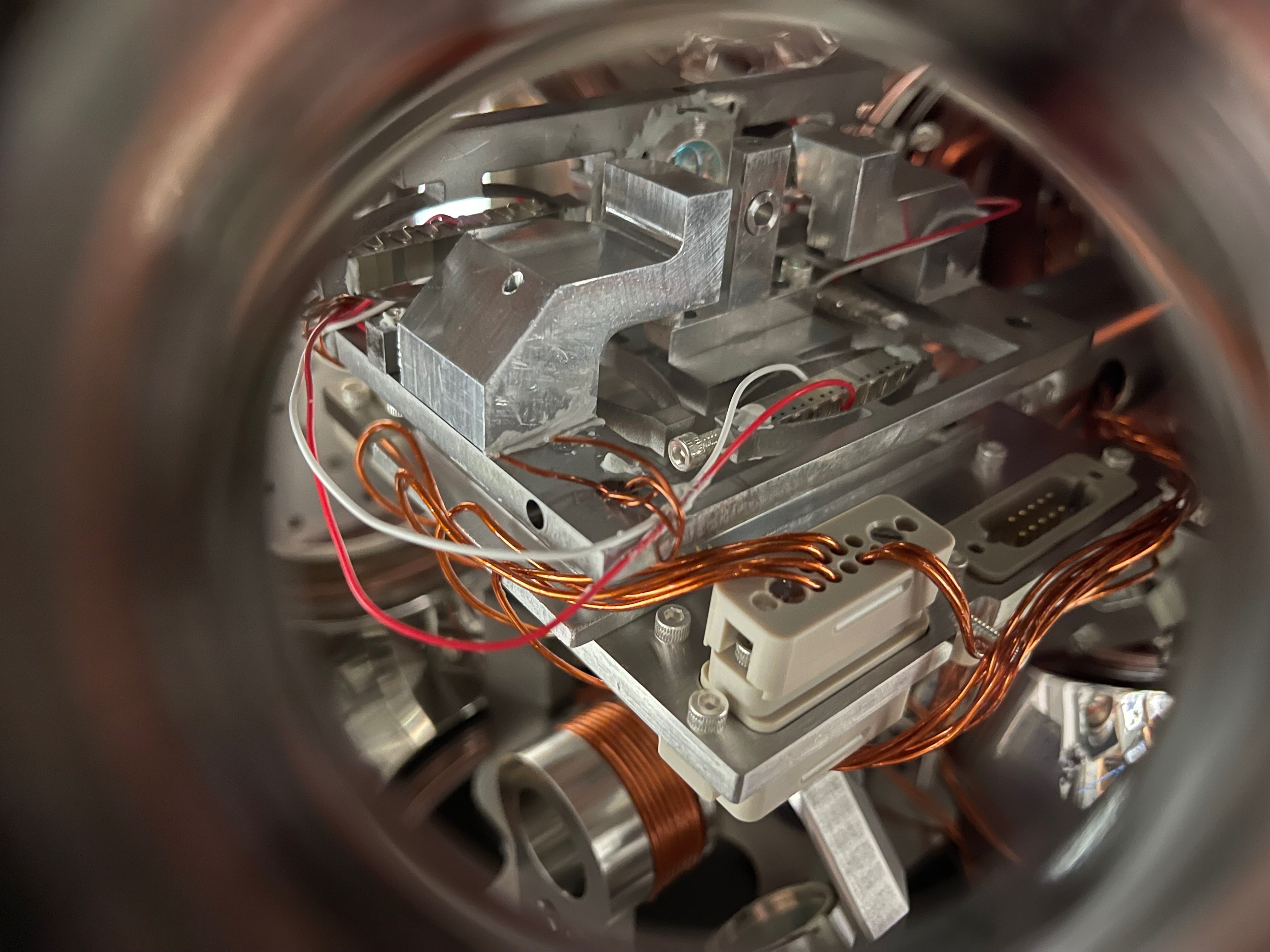}
	\caption{
		In vacuum installation of the completed flexure assembly, with aspheric lens and curved mirror glued into mounts.
	}
	\label{fig:flexure}
\end{figure*} 

As covered in Supplement~\ref{SI:StabilityDiagrams}, the difficulty in operating cavities at the small waist points of the stability diagram arises from the increased sensitivity of the cavity axis to transverse misalignments. This sensitivity is much improved for the small waist cavity geometry outlined in this paper. Nonetheless, precise in-vacuum positioning is still necessary to account for drifts in cavity alignment during the bake and to tune the relative position of optics to the stable point. Our custom designed/machined flexure mounts allow us to accomplish this task without the use of commercial integrated three-axis piezo stages, which can have long lead times, less throw, and high cost.

The cavity mount uses two waterjet-cut flexures for 3D control of the relative transverse alignment between the C140 asphere and the ROC = 1 cm curved mirror: 
\begin{enumerate}
    \item A single-axis vertical flexure mount for the curved mirror which provides control over the $y$-axis control
    \item A two-axis horizontal flexure that attaches to the asphere mount and provides $x$/$z$ motion
\end{enumerate}

Both flexures employ a similar lever arm design to amplify the base $45~\mu$m throw of the NAC2003-H32 Noliac piezo stacks. Initial designs used a straight lever arm, for which the theoretical amplification is given simply by the constrained ratio of displacements in piezo, set by the angle of the arm. Space constraints motivated switching to the curved metal flexing elements shown in Figure \ref{fig:MolassesMounts}. To simulate the predicted amplification for this design, we used Fusion's stress analysis software with factors included to account for the stress/strain parameters (Young's modulus, etc.) of stainless steel. For the vertical mount, two piezos were used (one for each side to preserve symmetry), and we use a Mach-Zehnder interferometer to measure an amplified throw of $\sim 96~\mu$m. The horizontal flexure operates in 2D, with mobile central platform supported by four lever arm slots. Only two of these four slots are mounted with piezo stacks -- one for control in $x$ and the other for $z$. We measure and amplified throw of $80~\mu$m in both directions.

The completed structure is secured using vented silver screws, with the vertical flexure sitting upright on the horizontal flexure. The entire assembly sits on top of a base plate to properly align the component relative to the loadlock translator arm. We use silver prism mirrors for the 3D molasses beams. These slide into custom machined molasses mounts (see Figure~\ref{fig:MolassesMounts}) with angles set to achieve cooling at the small waist.



\section{Model-Free Estimation of the Atom Detection Efficiency}
\label{SI:DetEfficelModelFree}

The figures of merit for assessing the performance of the cavity-enhanced readout demonstrated in this paper are (1) the atom detection fidelity for a measurement of time $t$ and (2) the survival rate of the atom during that measurement. To estimate these quantities, it is customary to tailor individual fitting routines to the separate sections of the single-atom histogram (background peak, atom peak, signal tails, etc). However, for fidelities above $99 \%$, the extrapolated fidelity becomes highly dependent on the functional form of the histogram in its low probability wings~\cite{madjarov2021entangling}. In a cavity the dependence of $g$ on the axial position of the atom further complicate this choice of functional forms by introducing averaging of the dipole signal over the nodes and anti-nodes of the readout mode, or by creating an axial variation in coupling strength beyond the Raleigh range, relevant for lattice signals where the signal atom can in principle be localized one of many sites.

To circumvent this issue we employ a model-free estimation scheme for calculating the atom detection efficiencies, as detailed in Section 2.6.7 of reference~\cite{madjarov2021entangling}. The method relies on extrapolating the survival rate S and fidelity F from the set of probabilities $p_{x_{1} x_{2}}$, where $x_{1}$ is 0 if the the signal is measured above the threshold in the first of two consecutive shots, and $x_{2}$ defined similarly for the the second shot in the measurement. The probabilities $p_{x_{1} x_{2}}$ can than be enumerated individually and related to F and S in a system of equations: 

\begin{equation}
    p_{11} = f F_{1}^2 S + (1-f)(1-F_0)^2 + f F_{1} (1 - S) (1 - F_{0})
\end{equation}

\begin{equation}
    p_{10} = f F_{1} S (1 - F_{1}) + f F_{1}(1-S) F_{0} + (1 - f) (1 - F_{0}) F_{0}
\end{equation}

\begin{equation}
    p_{01} = f (1 - F_{1}) S F_{1} + f (1 - F_{1})(1-S) (1 - F_{0}) + (1 - f) F_{0} (1 - F_{0})
\end{equation}

where $f$ is the filling fraction, and the overall fidelity $F = f F_{1} + (1 - f) F_{0}$. From the two shot data, it is  easy to extrapolate $f$, $p_{11}$, $p_{10}$, and $p_{01}$ at a given threshold and then to extrapolate F and S from the equation above. Generally, there will be two solutions, one of which can be discarded since as it will predict F < 0.5. For fidelities up to $99.9 \%$, the final F is robust to fluctuations in the loading fraction $f$, though the component fidelities $F_{0}$ and $F_{1}$ can be highly sensitive to it. There is only a narrow range of the loading fraction for which all solved parameters are physical, providing a check on the reasonableness of the extrapolated $f$.

\section{Comparing Cooperativity Expressions}
\label{SI:CooperativityExpressions}
Here we begin with the most commonly used expression for cooperativity, $C\equiv\frac{4 g^2}{\kappa\Gamma}$, and show that, up to numerical factors, it is equivalent to (a) the closed-transition cooperativity $C_{closed}=\frac{24F}{\pi}\frac{1}{(k w_0)^2}$, and (b) the Purcell factor, $F_P=\frac{3}{4\pi^2}\lambda^3\frac{Q}{V}$.

To show that $C=C_{closed}$, we note that for a closed transition, the atomic linewidth can be written using the Weisskopf-Wigner formula: $\Gamma\equiv \frac{\omega^3 d^2}{3\pi\epsilon_0\hbar c^3}$, where $d$ is the dipole moment of the transition. The cavity linewidth can be written in terms of the cavity length $L$ and finesse $F$ according to $\kappa=2\pi\times\frac{c}{2L}\times{1}{F}$. Finally, the vacuum Rabi coupling $g$ can be written as $g=d E_1/\hbar$, where $E_1$ is the electric field of a single photon in the cavity at the location of the atom. By conservation of energy $\frac{1}{2}\epsilon_0 E_1^2\times \frac{\pi}{2} w_0^2 L=\frac{1}{2}\hbar\omega$, so $g^2=\frac{2 d^2\omega}{\hbar\epsilon_0\pi w_0^2 L}$. Combining all of these results yields: $C\equiv\frac{4 g^2}{\kappa\Gamma}=\frac{24F}{\pi}\times\frac{1}{(k w_0)^2}=C_{closed}$, as anticipated.

To show the equivalence of the Purcell factor $F_P$, we just note that for a standing wave resonator of length $L$ and finesse $F$, the quality factor is given by $Q=\frac{F (2L)}{\lambda}$, and the mode volume by $V=\frac{1}{2}\frac{\pi}{2}w_0^2 L$. Combining these expressions yields $F_p=\frac{12 F}{2\pi^3}\frac{\lambda^2}{w_0^2}=\frac{24 F}{\pi}\frac{1}{(k w_0)^2}$, equal to $C$,$C_{closed}$.

In total, we can understand this equivalence to mean that, at least for atoms with closed transitions, the probability that the atom emits into a cavity is entirely independent of both transition's dipole moment, and length of the cavity. Atoms with narrow transitions (small dipole moments) emit slowly into cavities, but they emit into freespace \emph{even more slowly}. Similarly, a long cavity has a small $g$, and hence a small light-matter coupling strength, but it also has a narrow linewidth, so the weak coupling has more time to coherently reinforce itself.

The only caveat to this story is that weak transitions and long cavities result in slower atom emission into the resonator, and hence \emph{slower} readout; if readout speed is truly the currency, then choosing shorter cavities and stronger transitions is sensible.

\section{Photon Loss Budget}
\label{SI:LossBudget}

The speed with which high-fidelity single atom detection can be performed is set by the photon collection rate at the photodetector, accounting for all loss channels in the system. We enumerate these for our system at the high outcoupling (20$\%$) point:

\begin{enumerate}
\item \textbf{Cavity Collection: }15$\%$. At optimal Pellicle outcoupling, the cavity collects 15$\%$ of photons (see SI~\ref{SI:outcoupling}). This is double the free space collection efficiency 7.8 $\%$ of the ideal 0.56 NA of the C140. Our expression accounts for the birefringence of the cavity due to the EOM and Pellicle beamsplitter and for the delocalization due to the finite temperature of the atom. Without the birefringence the predicted collection efficiency increases to 25 $\%$, and for an atom perfectly localized to an anti-node this further rises to 35 $\%$.

\item \textbf{Fiber Coupling: }50~$\%$. Light that leaks out of the cavity on both sides of the Pellicle beamsplitter propagates past an 808 nm line filter, where locking and signal paths are divided, and through to the fiber incoupler. The 780 nm reflection of the LL01-808 Max Line filter from AVR optics can be optimized to be effectively loss (R > 99.9 $\%$). We measure the fiber coupling efficiency of both paths to be $\approx{50}\%$.

\item \textbf{Quantum Efficiency: }64~$\%$. We use single photon counter module SPCM-AQR-14, with a quantum efficiency of 64~$\%$ at 780 nm.

\end{enumerate}

Overall, we calculate a total collection efficiency of 4.8 $\%$. These parameters are far from optimized. With a more sophisticated locking and outcoupling scheme to obviate the need for birefringent intracavity optics together with the use of an EMCCD which does not require fiber-coupling and has a higher overall quantum efficiency (80 $\%$ at 780 nm), a fourfold increase in total collection efficiency to 20~$\%$ should be achievable. Assuming that detection time goes linearly with improvements in efficiency, this will enable readout with 99.5 $\%$ fidelity in 27 $\mu$s.  Beyond these improvements, the use of lenses with better anti-reflection coatings should enable further gains in the photon collection rate of the cavity by increasing the finesse and therefore the Purcell factor. At present, we extrapolate an internal loss of the cavity of $\approx{12 \%}$, based on our measured finesse at 4$\%$ outcoupling. This loss includes the relatively poor anti-reflection coating on the C140-TMD-B asphere with round trip loss of 1.2 $\%$ in addition to loss due to clipping. Both issues can be circumvented with better polished aspheres. This will enable high-fidelity detection in 10 $\mu$s.

\clearpage
\putbib
\end{bibunit}

\
\end{document}